\documentclass[twocolumn,prl,aps]{revtex4}

\usepackage{dcolumn}
\usepackage{bm}
\input{epsf.tex}

\begin{document}

\preprint{APS/123-QED}

\title{Quasiparticle spectrum of the cuprate Bi$_2$Sr$_2$CaCu$_2$O$_{8+\delta}$ \,:
\\ Possible connection to the phase diagram}

\author{W.\,Sacks$^1$, T.\,Cren$^1$, D.\,Roditchev$^1$ and B.\,Dou\c{c}ot$^2$}
\affiliation{$^1$Institut des NanoSciences de Paris, I.N.S.P.,
\\ Universit\'es Paris 6 et 7, C.N.R.S. (UMR 75\ 88), 75015 Paris,
France}

\affiliation{$^2$Laboratoire de Physique Theorique et Hautes
\'{E}nergies, L.P.T.H.E., Universit\'es Paris 6 et 7, C.N.R.S. (UMR
75\ 89), 75005 Paris, France}

\date{\today}

\begin{abstract}
We previously introduced [T. Cren et al., Europhys. Lett. {\bf 52},
203 (2000)] an energy-dependant gap function, $\Delta(E)$, that fits
the unusual shape of the quasiparticle (QP) spectrum for both
BiSrCaCuO and YBaCuO. A simple anti-resonance in $\Delta(E)$
accounts for the pronounced QP peaks in the density of states, at an
energy $\Delta_p$, and the dip feature at a higher energy,
$E_{dip}$. Here we go a step further\,: our gap function is
consistent with the ($T,\,p$) phase diagram, where $p$ is the
carrier density. For large QP energies ($E \gg \Delta_p$), the {\it
total spectral gap} is $\Delta(E) \simeq \Delta_p + \Delta_\varphi$,
where $\Delta_\varphi$ is tied to the condensation energy. From the
available data, a simple $p$-dependance of $\Delta_p$ and
$\Delta_\varphi$ is found, in particular $\Delta_\varphi(p) \simeq
2.3\, k_B\, T_c(p)$. These two distinct energy scales of the
superconducting state are interpreted by comparing with the normal
and {\it pseudogap} states. The various forms of the QP density of
states, as well as the spectral function $A({\bf k},E)$, are
discussed.

\end{abstract}


\maketitle

\centerline{\bf \small INTRODUCTION}

\vskip 4 mm

A striking feature of the conventional superconducting (SC) state
is the small number of parameters needed for its description. With
the knowledge of the BCS quasiparticle spectrum, $E_{\bf k} =
\sqrt{\epsilon_k^2 + \Delta_{\bf k}(T)^2}$, revealing a gap
$\Delta_{\bf k}(T)$ at the Fermi level ($\epsilon_k$=0), the
Hamiltonian is basically known and the magnetic, thermodynamic and
transport properties can be derived \cite{bcs}. The gap
$\Delta_{\bf k}(T)$, which vanishes at $T_c$, is related to the
pairing interaction $V_{{\bf k},{\bf k}'}$ via the BCS
self-consistency relation, giving the ratio $2 \Delta(0)/k_B T_c =
3.52$, in the weak-coupling isotropic case. It is thus a scalar
order parameter of the transition, a fact that has been verified
to a high precision \cite{parks,schrieffer,giaever}.

In the case of high-T$_c$, the probing of the QP spectrum has not
led to a solution. Still, a wealth of information on the magnetic
field, temperature and doping dependence of the SC state has been
obtained \cite{reviews,tallon}. The QP spectral function, as probed
using photoemission (ARPES), or the density of states (DOS) as
obtained by Scanning Tunneling Spectroscopy (STS), reveal additional
singularities which are at odds with a simple BCS $d$-wave spectrum
\cite{norman2, Campuzano, miyakawa, rennerprb, phd, constraints,
zaza1}. In the DOS (Fig.\,1, curve 1), the QP peaks ($P$) are very
pronounced and are followed by a dip feature ($D$) at higher energy
(at $E=E_{dip}$). Although the origin is still debated, there is
some strong-coupling effect on the quasiparticles : a self-energy is
implied \cite{coffey, norman3, pines, eschrig, millis}. A mean-field
approach is insufficient in the context of correlated electrons
\cite{anderson, millis2}, coupled spin-charge degrees of freedom
\cite{atkinson2, abanovSF, ohkawa}, phase fluctuations \cite{emery,
kwon, loktev}, or a competing order \cite{sachdev, chakravarty,
levin1, varma}. Thus the SC state can no longer depend on one
parameter.

The main question addressed in this work is can the QP spectrum
still be described in simple terms (eg. in an extended BCS way)
and if so, how does it reflect the order parameter\,? To answer,
the details of the quasiparticle DOS must be understood. As we
showed in ref.\,\cite{constraints}, the particular shape of the
measured spectrum, illustrated by curve 1, cannot be obtained from
a simple mean-field gap, giving curve 3. However, our resonant gap
function $\Delta(E)$, described further in Sec.\,I, nicely fits
the variety of spectra published since
\cite{phd,zaza1,zaza2,zaza3,hoogenboom,howald,pan,
sugimoto,wang,xuan,matsuba,kapit}. Despite a number of analyses of
the QP spectra, taking into account the coupling to a collective
mode \cite{abanov, zaza2, zaza3, abanovSF, abanovCP, chubukov,
bang, eschrig}, the effect of van Hove singularities or the
particle-hole asymmetry \cite{abrikosov, yusof, hoogenboom,
hirsch}, the effects of disorder or phase fluctuations
\cite{atkinson2, millis}, the spectrum 1 is difficult to derive.

\vskip 4mm

\begin{figure}[h]
\centering
    \vbox to 4.6 cm{
    \epsfxsize=7.2cm
    \epsfbox{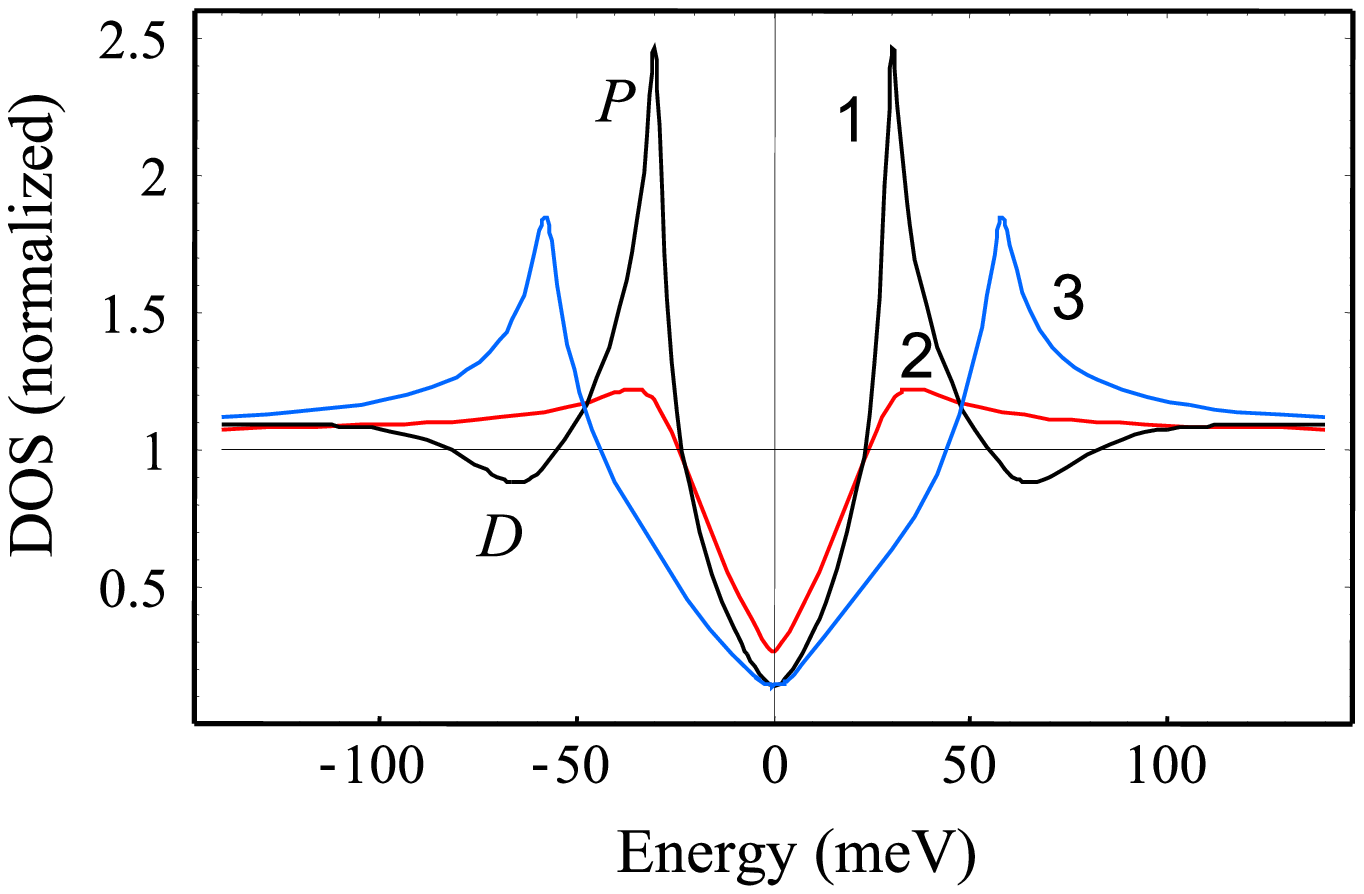}
    }
    \end{figure}

{\small Fig.\,1\,: Curve 1 - Quasiparticle DOS as observed by
tunneling showing pronounced peaks at $E=\pm\Delta_p$ followed by
dips. Curve 2 - {\it Pseudogap} type spectrum observed in the
vortex core, or for $T \geq T_c$. Curve 3 - Extended BCS
($d$-wave) DOS, but with a larger gap. All three are generated by
our gap function, Eq.\,(10).}

\vskip .2 cm

The physical parameters of a self-energy, or equivalent gap
function, have yet to be connected to the phase diagram. Important
leads on the spectral function (or DOS), as a function of doping
and temperature, have been obtained using ARPES and tunneling.
First, a well-defined QP peak at the position $\Delta_p$ (our
notation) develops in the SC state and is of $d$-wave symmetry
\cite{vanh,ding2,ding1,loeser,norman}\,: $\Delta_{\bf k} \propto
\Delta_p\,{\rm cos}(k_x-k_y)$, yielding the characteristic V shape
in curves 1 - 3. However, the ratio $2 \Delta_p/k_B T_c$ widely
departs from the BCS value; $\Delta_p$ decreases roughly linearly
with $p$, with a large negative slope, from underdoped to
overdoped sides of the phase diagram \cite{miyakawa, ding2,
tallon}. As is well established, $T_c(p)$ is dome shaped with a
maximum at $p_0\simeq .16$, where $T_c \simeq 95$ K, for
Bi$_2$Sr$_2$CaCu$_2$O$_{8+\delta}$. Thus, the position of the QP
peak (at $E=\Delta_p$) is not the energy scale the global SC order
parameter.

The QP spectrum has been probed as a function of rising temperature
\cite{miyakawa, ding2, ding1, loeser, norman, kam, rennerT}. Above
$T_c$, instead of revealing the normal state with a metallic DOS,
the spectrum displays a pronounced but peak-less gap of width
2$\Delta_{PG}$ at the Fermi level. This {\it pseudogap}, see curve
2, disappears at the higher temperature $T^*$, and has possibly the
same angular dependence as $\Delta_{\bf k}$. Moreover, $\Delta_p$
and $\Delta_{PG}$ have approximately the same magnitude \cite{ding2,
kam, rennerT}. One finds that $\Delta_p\approx 3 k_B T^*$, so
$T^*(p)$ follows the identical trend as $\Delta_p$ in the phase
diagram. The challenge is to understand three contiguous phases
(superconducting, pseudogap and normal). However, the $T^*(p)$ curve
on the overdoped side, where the data is rare, is the subject of hot
debate \cite{tallon, millis2, levin1, loktev}, in particular whether
or not it crosses the $T_c(p)$ dome. The $T^*$ inferred from NMR
Knight shift, resistivity and specific-heat measurements may
correspond to still a higher temperature, such as the onset of
anti-ferromagnetic fluctuations, distinct from the vanishing of
$\Delta_{PG}$ as observed by ARPES \cite{comment}.

The origin of the pseudogap is a key question about high-T$_c$,
and is still highly controversial. Theories fall into two
qualitative categories\,: $\Delta_{PG}$ is either a precursor to,
or competes with, the SC state \cite{levin2}. The first
\cite{emery,levin2,loktev} is immediately compelling due to the
phase diagram\,: below $T^*$ incoherent pairs are formed (thus the
pseudogap) and condense at $T_c$. It also follows many aspects of
the SC state (low carrier density, strong coupling energy, small
coherence length, sensitivity to disorder, etc.) and is
theoretically tractable \cite{emery, levin2, huscroft, ghosal}. In
the second category, the competing order (charge or spin density
wave, Varma currents, RVB,...\cite{sachdev, varma, chakravarty,
anderson}) sends $T^*$ down, possibly across the $T_c$ dome, to a
quantum critical point above which a new phase is formed. Deciding
between these two categories would be a significant advance.

Conductance mapping using STS has provided valuable information.
The `normal' state found within the vortex core \cite{phd,
rennerB} reveals a pseudogap analogous to the one found just above
$T_c$ \cite{rennerT}. The local vanishing of the SC order
parameter in a vortex core is due to the phase singularity; but
contrary to the conventional case, the pseudogap persists. We
found a quasi-identical pseudogap, in zero magnetic field, caused
by weak disorder \cite{prlmaps}. In both cases, $\Delta_p$ and
$\Delta_{PG}$ have about the same magnitude \cite{phd, rennerB,
matsuba, prlmaps, nanomaps}. Thus an important constraint on the
SC gap function is its smooth transition to the peak-less
pseudogap when phase coherence is lost\,: curve 1 $\rightarrow$
curve\,2 (Fig.\,1).

\vskip 2 mm

Nozi\`{e}res and Pistolesi \cite{pn} have described a superconducting
state when a precursor gap, such as $\Delta_{PG}$, is initially
present. However, their model implies that the SC gap must be
larger than the precursor gap. Thus, given curve 2 for the
pseudogap, curve 3 is expected in the superconducting state, not
curve 1. Moreover, the high spectral weight of the QP peaks (curve
1) leads to an apparent paradox\,: electronic states seem to move
towards the Fermi level in the transition to the SC state
\cite{normanke}. Any complex self-energy or gap function must give
the correct energy change for the pseudogap to SC transition.

The situation becomes complex in the case of strong disorder\,:
local STS mapping has revealed pseudogap/SC gap variations at the
surface of BSCCO \cite{nanomaps, howald, pan, kapit}. Clearly, in
this case, there are changes of both SC amplitude and phase
\cite{atkinson, atkinson2, huscroft, hudson, balatsky}. In
Pb-substituted BSCCO the disorder causes `superconducting' islands
to form \cite{nanomaps}, where the intensity of the spectral fine
structure correlates with the degree of long-range order. Recent
theoretical work by Atkinson \cite{atkinson2} and experimental STS
by Fang et al. \cite{kapit} corroborate this conclusion. Even if
the inhomogeneity is not an intrinsic property
\cite{hoogenboom2,wang, tallon}, the cuprate SC state is sensitive
to local perturbations \cite{hudson, atkinson, balatsky}, and the
attenuation of the spectral fine structure is a clear sign. Then,
several parameters are needed for the interpretation of the
tunneling DOS.

In this Article, we analyze the QP spectrum, $E_{\bf k}$, as
inferred from the sharpest tunneling DOS of BSCCO. The detailed
DOS shape (pronounced QP peaks, followed by dips at higher energy)
is due to a single resonance in our energy-dependant gap function,
Section\,I. Such a resonance in the QP spectrum is possibly the
coupling to a collective mode \cite{abanov, abanovCP,zaza2, zaza3,
chubukov, norman3, eschrig} of the same origin as the resonance
seen using inelastic neutron scattering \cite{bourges} and which
scales with $T_c$\,: $\Omega \sim 5.3\, k_B\, T_c$. Zasadzinsky et
al. studied the dip position as a function of doping using
strong-coupling theory \cite{zaza1}, and suggested that the dip
energy is also related to $T_c$. Since the origin of the resonant
gap function (or QP self-energy) is still unknown, our aim is to
show how the basic parameters depend on the carrier density $p$.
In Sec.\,II we find that one of the energy terms,
$\Delta_{\varphi}$, is compatible with an order parameter\,: it is
proportional to $k_B\,T_c(p)$. The predicted shape of the QP
density of states is then studied as a function of $p$. We treat
the transition to the PG state (Sec.\,II), and the role of the two
distinct energy scales (Sec.\,III). Finally, the QP spectral
function and self-energy are discussed (Sec.\,IV).

\newpage


\section{I. Superconducting gap function}

Here the quasiparticle DOS, having the essential characteristics
of the observed STS conductance spectra, is derived. In Fig.\,2,
we show such a spectrum obtained on BSCCO, near optimally doped,
from our group (2a) which is compared to (2b), a spectrum from
Pan, Hudson et al. \cite{phd}. One can again see the pronounced QP
peaks, the steep slope on the outer side of each peak, followed by
the dip feature previously described. In this Section, we focus on
these main aspects of the DOS; the questions of the background
slope (as in Fig.\,2b), the Fermi surface anisotropy, the
particle-hole asymmetry and the van Hove singularity, have already
been given extensive treatment \cite{mallet, kitazawa,yusof,
hirsch,hoogenboom}. The detailed fits of Fig.\,2, solid lines,
give the key parameters of our SC gap function, without such
considerations.


\vskip 2 mm

{\it Expression for the QP-DOS}

\vskip .5 mm

Consider the spectral function, as measured by ARPES
\cite{Campuzano, ding2, ding1, norman, normanke, loeser, kam,
norman2}, in the two-dimensional model with $\vec k = (k, \theta)$
the wave vector in the $ab$ plane\,:
\begin{equation}
A({\bf k}, E) = \frac {1}{\pi}\ Im\ G({\bf k}, E)
\end{equation}
where $G({\bf k}, E)$ is the single-particle Green's function. The
superconducting DOS is then \cite{eschrig, propagator}\,:
$$ N_s(E) = \sum_{{\bf k}} \ A({\bf k}, E)$$
which is measured in the tunneling experiment. Converting sums to
integrals in the usual way\,:
\begin{equation}
N_s(E)  = \frac {N_n(0)}{2\pi}\ \int_0^{2\pi} d\theta\
\int_{-\infty}^{\infty} d\epsilon_k \ A({\bf k}, E)
\end{equation}
where $\epsilon_k$ and $N_n(0)$ are the normal excitation spectrum
and Fermi-level DOS, respectively. This expression ignores the
effect of the Fermi surface anisotropy \cite{mallet, yusof}. In
the case of an ideal quasiparticle with zero lifetime broadening,
and dispersion $E_{{\bf k}}=\sqrt{\epsilon_k^2 + \Delta_{\bf
k}^2}$, the propagator is\,\cite{propagator, schrieffer}:
$$
G({\bf k}, E) = \frac{u_k^2}{E-E_{\bf k} + i 0^-} +
\frac{v_k^2}{E+E_{\bf k}+ i 0^-}
$$
with $u_k, v_k$ the usual coherence factors. Then, $A({\bf k}, E)$
is just: $A({\bf k}, E) = u_k^2\,\delta(E- E_{\bf k}) + v_k^2
\,\delta(E + E_{\bf k})$ where each term corresponds to a
quasiparticle added ($E>0$) or removed ($E<0$), respectively.

For example, with fixed $E>0$, the integral in (2) picks up two
poles at $\pm \epsilon_k$ of amplitude $u_k^2(-\epsilon_k)$ and
$u_k^2(+\epsilon_k)$. As is well known \cite{schrieffer,
propagator}, the coherence factors disappear in the symmetric case,
since $u_k^2(-\epsilon_k) + u_k^2(+\epsilon_k) = 1$. We shall
thenceforth ignore them, and consider for both tunneling and ARPES
that the values $\pm \epsilon_k$ are equivalent. Using Eq. (2) the
expected result is obtained\,:
\begin{equation}
N_s(E)  = N_n(0)\ \frac {1}{2\pi}\ \int_0^{2\pi} d\theta\
\left(\frac{\partial \epsilon_k}{\partial E_{{\bf
k}}}\right)_{E_{\bf k}=E}
\end{equation}
It is then convenient to define the partial (one dimensional) DOS at
the angle $\theta$\,:
\begin{equation}
N_s(E)  = \int_0^{2\pi} n_s(E, \theta)\ d\theta\
\end{equation}
and, in the extended BCS $d$-wave model \cite{mallet, maki}, the
partial DOS is\,:
\begin{equation}
n_s(E, \theta) = \frac{N_n(0)}{2\pi}\
\frac{E}{\sqrt{E^2-\Delta_p^2\,{\rm cos}^2(2\theta)}}
\end{equation}
Expressions (4, 5) can be used to generate the curve 3 in Fig.\,1
which, as we have stressed, fails to match the strong spectral
weight of the QP peaks, and the dip features, obvious in the
experiments (Fig.\,2).

\begin{figure}[h]
\centering
    \vbox to 8.6 cm{
    \epsfxsize=10.5cm
    \ \ \ \epsfbox{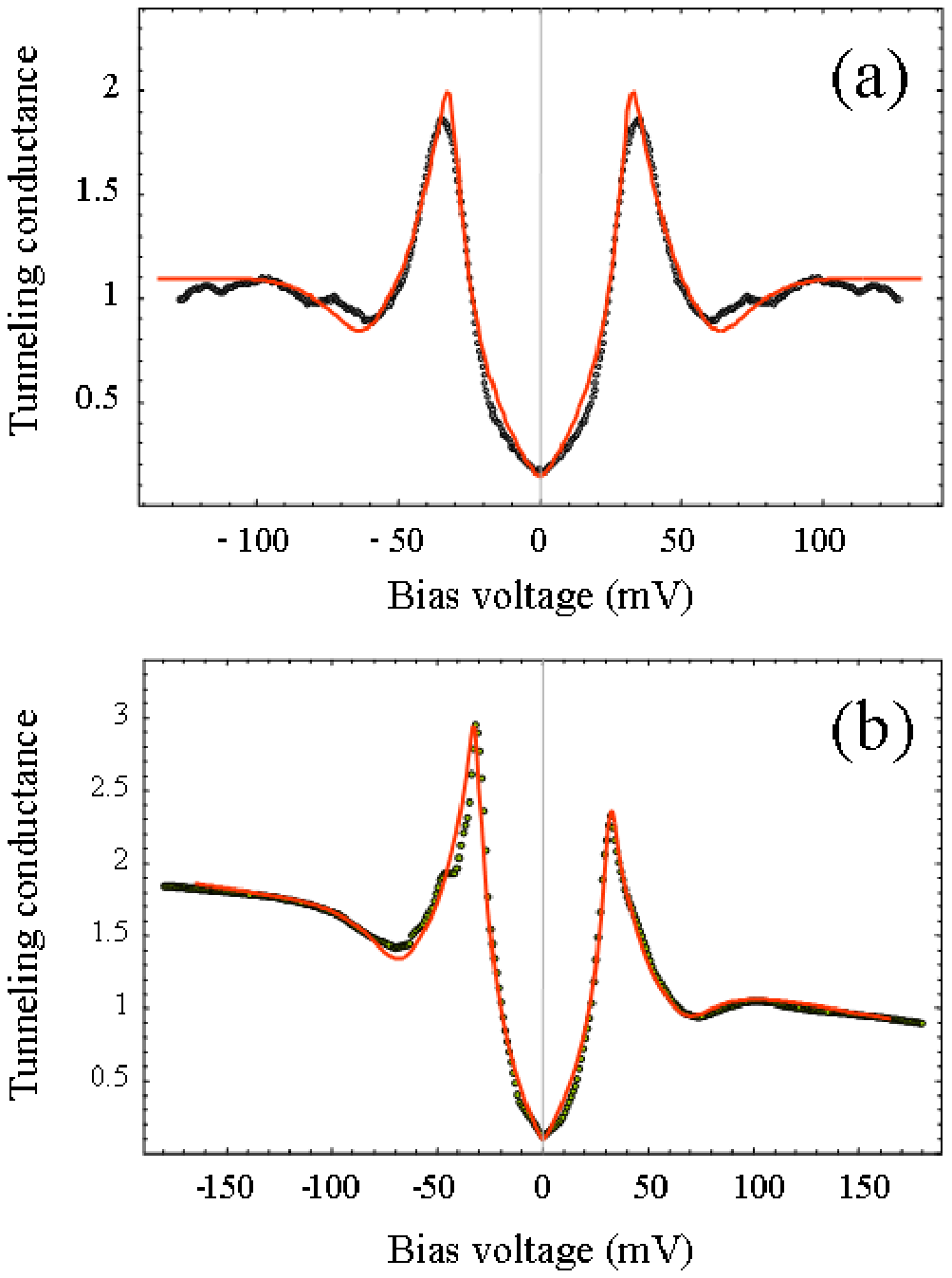}
    }
    \end{figure}
{\small Fig.\,2 : Dotted lines - local STS conductance spectra.
Solid lines - fits using the model (9, 10).

(a) Symmetrized spectrum from ref.\cite{constraints}, with
background removed. Here $\Delta_p=31.5$ and $E_{dip}=64$ are
fixed; the two free parameters are\,: $\Delta_\varphi = 18.5$ and
$E_0 = 50$, all in meV. Also, $E_0 = \Delta_0\,$,
$\eta=\Delta_\varphi$ and $A$=2, as seen in Fig.\,3. The
broadening values are $\Gamma=.08\,\Delta_p$ and $\delta = 0.1$.

(b) Spectrum from Pan et al.\,\cite{phd}. The gap function (10) is
used but the parameters, $\Delta_p(p)$ and $\Delta_\varphi(p)$,
are from the $p$-dependent fit to the data (Fig.\,4a). The best
fit is for $p=.175$, with the background slope added. All
parameters are nearly identical with (a) except $E_0\simeq 1.06
\,\Delta_0\,$ and $\Gamma=.03\,\Delta_p$. }

\vskip 2 mm

In the general case, beyond the mean-field approach, one must
evaluate\,:
\begin{equation}
n_s(E, \theta) = \frac {N_n(0)}{2\pi}\  \int d\epsilon_k \ A({\bf
k}, E)
\end{equation}
assuming a suitable model for $A({\bf k}, E)$ \cite{coffey,
abanovSF, abanovCP, millis, zaza3, eschrig}. Our approach is to
consider that strong-coupling modifies the electron-electron
interaction, but {\it without the retarding effects} that would
occur in the case of phonon-mediated pairing \cite{constraints}.
Thus we write $\Delta_{\bf k} \rightarrow \Delta_{\bf k}(E_{\bf
k})$, and the new dispersion law is\,:
\begin{equation}
E_{\bf k} = \sqrt{\epsilon_k^2 + \Delta_{\bf k}(E_{\bf k})^2} +
i\,\Gamma
\end{equation}
where we add $\Gamma$, the lifetime broadening introduced by Dynes
\cite{dynes}. Assuming particle-hole symmetry, we use\,:
\begin{equation}
A({\bf k}, E) = \frac {1}{\pi}\ Im\ \frac{1}{E-E_{\bf k}+ i 0^-}
\end{equation}
and, performing the complex integration in (6), we obtain for the
partial DOS\,:
\begin{equation}
n_s(E, \theta) =  \frac {N_n(0)}{2 \pi}\ Re\ \frac{ E - i \Gamma -
\Delta_{\bf k}(E) \frac{\partial \Delta_{\bf k}(E)}{\partial E} }
{\sqrt{ (E - i \Gamma)^2 - \Delta_{\bf k}(E)^2}}
\end{equation}

Here $\Delta_{\bf k}$, evaluated at the pole ($E_{\bf k}=E$), is a
function of $\theta$. For a constant gap $\Delta_{\bf k} =
\Delta_p$, we get back Dynes's formula for the BCS DOS with the
lifetime broadening. In the general case, (9) is the basic
equation for the quasiparticle DOS in our approach, once
integrated over $\theta$. It contains a new term in the numerator,
$ -\Delta_{\bf k}\partial \Delta_{\bf k} /\partial E$ which is
responsible, as already shown in \cite{constraints}, for the
distinct modification of the DOS seen in Fig.\,2. It can be used
to match the tunneling data of \cite{rennerprb, constraints, wang,
sugimoto, phd, pan, howald, matsuba, kitazawa, miyakawa, mallet,
zaza1, prlmaps, xuan, kapit, nanomaps}.

\vskip 2 mm

{\it Superconducting gap function}

\vskip 1 mm

We now consider the particular gap function appropriate for fitting
the data (solid lines in Fig.\,2). For the superconducting state, we
have made a slight change in notation with respect to ref.
\cite{constraints}.  Assuming $\Delta_{\bf k}(E) = {\rm
cos}(2\theta)\Delta(E)$, the gap function along the anti-nodal
direction is now written\,:
\begin{equation}
\Delta(E) = \Delta_p + \Delta_\varphi (1 - g(E))
\end{equation}
where $\Delta_p$ and $\Delta_\varphi$ are constant parameters and
$g(E)$ is a simple Lorentzian\,:
\begin{equation}
g(E) = A\ \frac{\eta^2}{(E-E_0)^2+\eta^2}
\end{equation}
having the standard parameters. Thus the second term in (10) is an
anti-resonance (a local decrease in the pair potential). It provokes
an additional peak, and dip, in the DOS near the two possible
extrema of $dg(E)/dE$ (see upper panel, Fig.\,3).

Consider the problem of fitting the DOS in some systematic way\,:
there are {\it a priori} 5 parameters (the role of $\Gamma$ will
be discussed subsequently). We commence with the data of Fig.\,2a,
where the spectrum is symmetrized and the background removed. From
our previous work \cite{constraints}, the resonance energy must
lie between the QP peaks and the dip position; an estimate is $E_0
\sim (\Delta_p+E_{dip})/2$. This condition ensures that the QP
peaks are reinforced, where the extremum of $-d\Delta/dE $ is
positive, and gives the dip at a higher energy, where $-d\Delta/dE
$ is a negative extremum (see Fig.\,3).

\begin{figure}[h]
\centering
    \vbox to 9.2 cm{
    \epsfxsize=8.6 cm
   \epsfbox{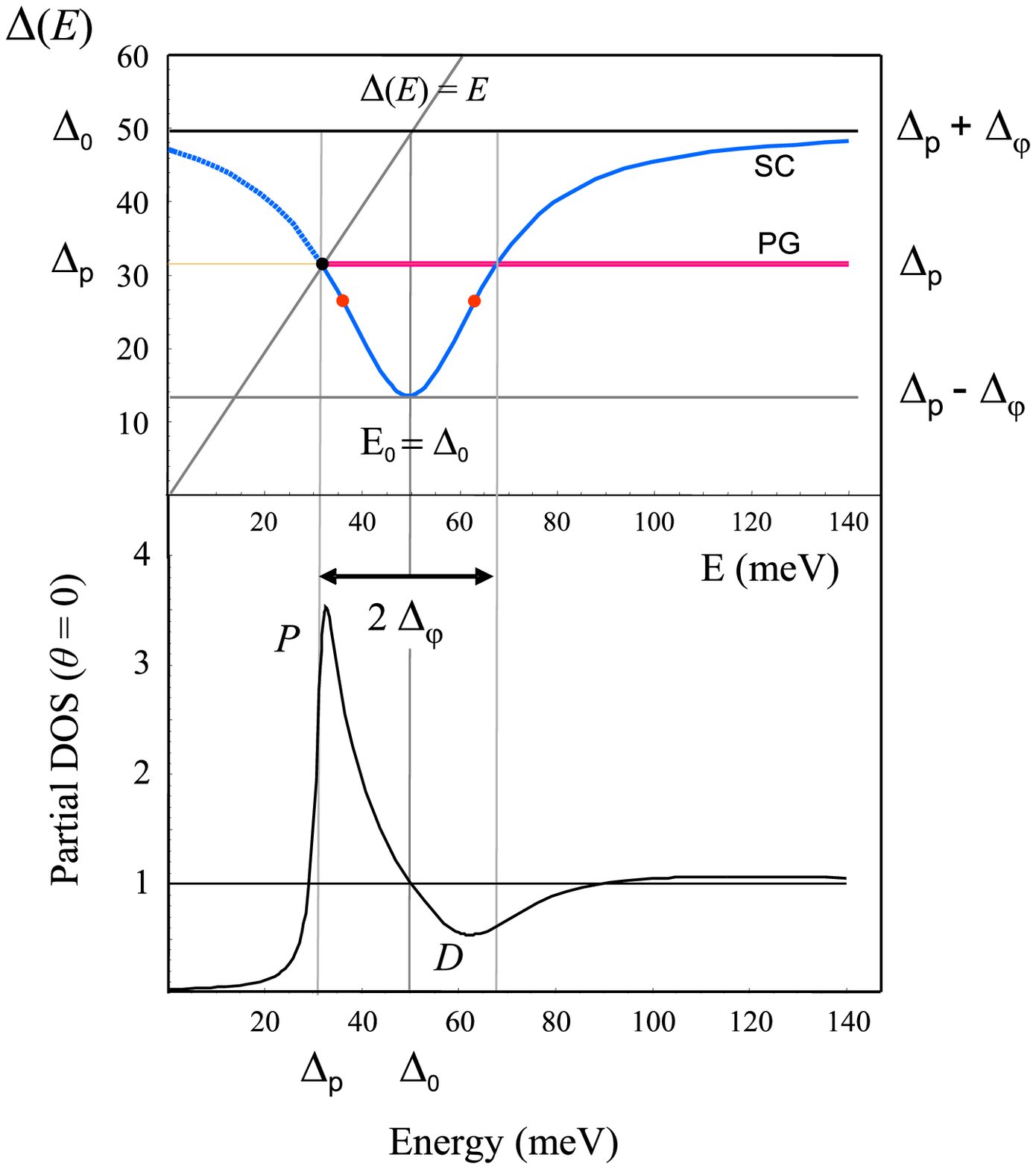}
    }
    \end{figure}
{\small Fig.\,3 : Characteristics of the SC gap function and their
link to the DOS along the anti-nodal direction ($\theta=0$).

Upper panel - Real part of $\Delta(E)$ from the fit of spectrum
Fig.\,2a. Note the anti-resonant shape, with the minimum at the
energy $E_0=\Delta_0$, while both amplitude and width are
$2\,\Delta_\varphi$. The intersection point, along $\Delta(E) =
E$, gives the QP peak at $\Delta_p$. For large energies, the
pairing interaction, is $\Delta(E) \simeq \Delta_0 = \Delta_p +
\Delta_\varphi$.

Lower panel - Partial DOS showing a sharp QP peak (P), followed by
the dip (D) at the higher energy $E_{dip}$. The {\it derivative}
of the gap function, as in Eq.\,(9), reinforces the quasiparticle
peak (negative extremum) and causes the dip feature (positive
extremum).}

\vskip 2 mm

Since $\Delta_p $ is {\it defined} as the QP peak position, easily
estimated from the data, we have the further condition that\,:
$\Delta(\Delta_p)= \Delta_p$, also illustrated in Fig.\,3. Using
(10) and (11), this gives $g(\Delta_p)=1$, or\,:
\begin{equation}
A = 1 + \frac{(E_0-\Delta_p)^2}{\eta^2}
\end{equation}
as a constraint on the parameters. A second constraint is found by
writing the dip position, $E_{dip}$, using the analytical
expression of the DOS. Considering the extremum of $\Delta
d\Delta/dE$ leads to the good approximation\,: $E_{dip}\simeq E_0
+ .73\,\eta$. Therefore, taking $E_{dip}$ and $\Delta_p$ to be
known, the fit to the spectrum of Fig.\,2a can be done by the
variation of only two free parameters\,: $E_0$ and
$\Delta_\varphi$. Their final values determine the precise
concavity in the DOS, between the QP peak energy and $E_{dip}$,
which is {\it a priori} unknown.

For the fit of spectrum 2a, using $\Delta_p=31.5$ and $E_{dip}=64$,
we obtain the values $E_0 = 50$, and $\Delta_\varphi = 18.5$, all in
meV. The problem is thus simplified by the following outcome\,: $E_0
= \Delta_p + \Delta_\varphi$ = $\Delta_0$, with
$\eta=\Delta_\varphi$ and $A$=2.

The corresponding function $\Delta(E)$ is depicted in the upper
panel of Fig.\,3, with the partial DOS along $\theta=0$ for direct
comparison (lower panel). One can see that $\Delta(E)$ has a
minimum at $E_0=\Delta_0$, where its value is
$\Delta_p-\Delta_\varphi$, then it increases towards the
asymptotic value $\Delta_p+\Delta_\varphi$, when $E \gg E_{dip}$.
Note that the total amplitude of the resonance, $2
\Delta_\varphi$, is identical to its width. In short, we obtain\,:
$$\Delta_0-\Delta_p=E_0-\Delta_p=\eta=\Delta_\varphi \ ,$$and we
propose that these relations should scale when the carrier
density, $p$, varies (Section II). The dip position, near to
$\Delta_p+2\Delta_\varphi$, is more precisely\,:
$$ E_{dip}\simeq
\Delta_0 + .73\,\Delta_\varphi \simeq \Delta_p +
1.73\,\Delta_\varphi
$$
so that $E_{dip}-\Delta_p \propto \Delta_\varphi$. The latter
parameter thus plays a fundamental role in our model.

\vskip 2 mm

{\it Broadening parameters}

\vskip 1 mm

Two broadening parameters in the final fits of Fig.\,2 are used.
First, the $i\,\Gamma$ introduced by Dynes to treat a finite
quasiparticle lifetime, artificially displaces the pole in the
spectral function (8) off the real axis. It is equivalent to a
Lorentzian broadening of the DOS, of full-width $\sim 2\Gamma$,
affecting the states at all energies. Consequently, in the lower
panel of Fig.\,3, virtual states lie within the gap of the partial
DOS (for $E<\Delta_p$). As we discuss in a recent paper
\cite{bergeal}, several factors can contribute to $\Gamma$ both
intrinsic (inelastic scattering, many-body effects,...) and
extrinsic (high-frequency noise,...). In Fig.\,2, the $\Gamma$ value
was adjusted to fit the zero-bias conductance.

The QP peaks, as calculated using (10), are initially higher than
those of Fig.\,2. Since $\Gamma$ is fixed, a second broadening
parameter is introduced\,: we replace $\Delta_p$ by a complex number
with a small imaginary part, $\Delta_p \rightarrow \Delta_p (1 - i
\delta)$. This has a major effect on the QP peaks, but a small one
on the remaining spectrum. Intuitively, the imaginary part
represents a smearing of the value of $\Delta_p$, such as in the
case of gap anisotropy \cite{bergeal} or Doppler shifts due to
supercurrents \cite{kohen}. While this ansatz is used to model the
pseudogap in Section II, it has a negligible influence on the fit
parameters deduced here.

\section{II. Connection to the phase diagram}

The fit to the BSCCO spectrum, Fig.\,2a, leads to the simple
result that the width, amplitude and position of the
anti-resonance are all simply related to the quantity
$\Delta_\varphi$; assuming $\Delta_p$ to be known. The QP spectrum
therefore depends on only {\it two} energy scales. This section is
devoted to their possible link to the phase diagram\,: we infer
how the parameters of the SC gap function change with the carrier
density.

\vskip 2 mm

{\it Doping dependance of the energies}

\vskip 1 mm

The foregoing suggests that $\Delta_\varphi$ must have a new
meaning. Consider the asymptotic value of $\Delta(E)$ for large
$E$\,: $\Delta(E)\simeq \Delta_p + \Delta_\varphi = \Delta_0$ for
$E \gg E_{dip}$ (Fig.\,3). $\Delta(E)$ is then constant up to some
higher cut-off energy, as in BCS theory. Along the anti-nodal
direction\,: $E_{\vec k} \approx \sqrt{\epsilon_k^2 +
\Delta_0^2}$, and $\Delta_0$ is thus interpreted as the {\it total
spectral gap} in the SC state, even though the QP peak remains at
the smaller energy $\Delta_p$. The pseudogap in the vortex core
\cite{rennerB, phd, matsuba}, where phase coherence is lost, takes
on a value $\Delta_{PG} \simeq \Delta_p$ and, aside from the
thermal broadening, the same holds for the pseudogap just above
$T_c$ \cite{rennerT, ding2, kam}.

\vskip 1 mm

\begin{figure}[h]
\centering
    \vbox to 9.4 cm{
    \epsfxsize=8.0 cm
    \epsfbox{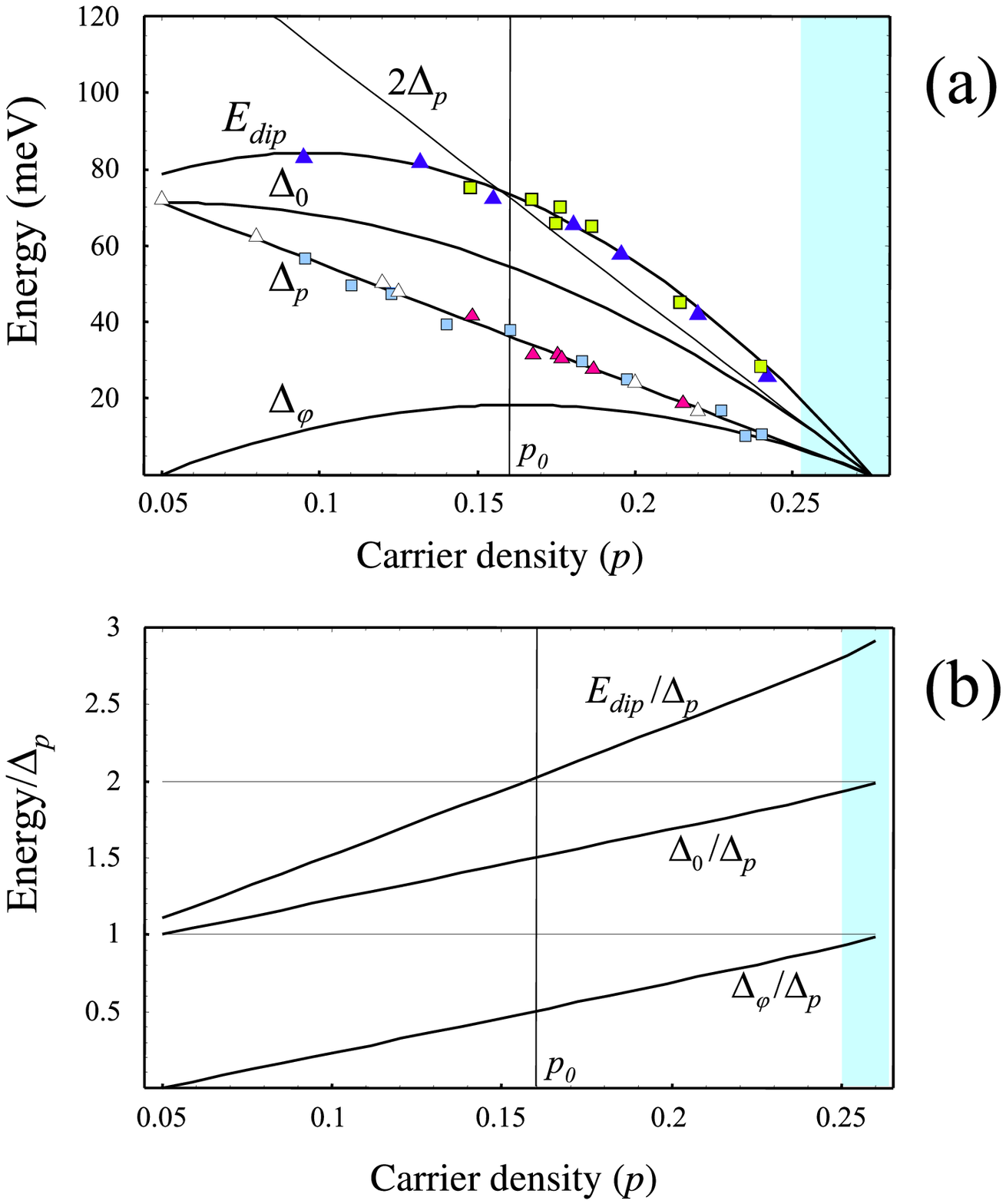}
    }
    \end{figure}

{\small Fig.\,4. Doping dependence of the two basic parameters of
the gap function, $\Delta_p(p)$ and $\Delta_\varphi(p)$, their
sum, $\Delta_0(p)$, and the dip position $E_{dip}$.

Upper panel (a) - $\Delta_p(p)$ line\,: Linear fit to the QP peak
positions taken from ARPES data \cite{ding2, Campuzano, kam},
SIS/SIN tunneling \cite{zaza1, zaza2, zaza3, miyakawa}, and STM
\cite{rennerprb, wang, sugimoto, phd, mallet, constraints}.
$E_{dip}$ line\,: Best quadratic fit to the dip position, as
estimated from STM data \cite{rennerprb, wang, sugimoto, phd,
mallet, constraints} and SIS tunneling \cite{zaza1, zaza3,
miyakawa}. The solution of the fit gives $\Delta_\varphi(p) =
2.3\,k_B\,T_c(p)$ and $E_0 = 1.06\,\Delta_0(p)$.

Lower panel (b) - Analogous diagram to (a) but with parameters
{\it normalized} to the gap energy $\Delta_p$\,: simple linear
laws are obtained, Eqs.\,(14-16). The principal energy scale of
the SC state, and determining the DOS fine-structure, is
$\Delta_\varphi(p)$. No evidence for a critical point is seen for
$0.05 < p < $0.25. }

\vskip 3 mm

We now suggest that $\Delta_\varphi$ is tied to the condensation
energy. The energy (per pair) of the SC state is then $\sim
-\Delta_\varphi$, with respect to the PG state, but it is $\sim
-\Delta_0$ with respect to the normal state. As will be discussed
in Sec.\,III, the integration over energy states, involving the
full $\Delta_{\bf k}(E_{\bf k})$, is necessary to obtain the
precise energy changes.

We can thus put\,: $\Delta_\varphi = C\, k_B T_c$ where $C$ is to
be determined. From the fit of Fig.\,2a, using $T_c \simeq 90$ K,
and $\Delta_\varphi = 18.5$ meV, gives $C \simeq 2.4$.

With this result, one could use scaling arguments to infer the
behavior of the parameters as a function of $p$. However, in view
of the dispersion of the tunneling and ARPES data, using a single
spectrum and one value of $T_c$ is restrictive. We take a
different approach by first plotting in Fig.\,4a a larger set of
data of the QP peak position, $\Delta_p$. Assuming that the
critical temperature $T_c(p)$ is known \cite{tallon}, one can
deduce the linear behavior of $\Delta_p(p)$ by a quadratic fit
\cite{miyakawa}. The data are from the SIS (SC-vacuum-SC) break
junction \cite{zaza1, zaza2, zaza3, miyakawa}, which probes the
electronic structure deep in the sample; SIN tunneling data, as
well as ARPES data, are included \cite{ding2, Campuzano, kam}. The
dip position ($E_{dip}$), as seen in many different tunneling
experiments \cite{rennerprb, wang, sugimoto, phd, mallet,
constraints, zaza1, miyakawa, zaza3}, is also plotted in Fig.\,4a.
The points evidently follow a continuous curve throughout the
phase diagram. We propose to find this curve via a best quadratic
fit using the gap function, which fixes all the parameters.

The two energy terms are thus $\Delta_p(p)$ and
$\Delta_\varphi(p)= C\, k_B T_c(p)$ where the best value of $C$ is
to be determined. In order to calculate $E_{dip}$ we could use the
previous condition $E_0 = \Delta_0$, which led to $E_{dip}\simeq
\Delta_p + 1.73\,\Delta_\varphi$. However, the spectrum Fig.\,2b,
from Pan et al., cannot be precisely fit with the resonance energy
exactly at $\Delta_0$. This is the essential difference between
the two spectra (2a and 2b); the latter dip position is slightly
at a higher energy.

From our study of YBCO in \cite{constraints}, we noticed the
resonance energy can be larger than $\Delta_0$. We thus write $E_0 =
\lambda \Delta_0$, in the general case, and determine the value of
$\lambda$ at the same time as $C$. The expression for the dip energy
is then\,:
\begin{equation}
E_{dip} = \lambda \Delta_0 + .73\, (\lambda \Delta_0- \Delta_p)
\end{equation}
and all other parameters retain their previous meaning. Putting
$\Delta_p(p)$ and $\Delta_\varphi(p)= C k_B T_c(p)$ in (13), and
fitting the data of Fig.\,4a, yields\,: $C=2.3$, $\lambda=1.06$,
and the continuous curve $E_{dip}(p)$. The two values differ by
about 5\% from the single-point estimate and, considering the
uncertainty in the data, the earlier observation in
\cite{constraints} that $E_0 \sim \Delta_0$ is maintained. More
significantly, we find that $\Delta_\varphi(p)= 2.3\, k_B T_c(p)$
is consistent throughout the phase diagram, from underdoped to
overdoped sides.

In Fig.\,4a the total spectral gap $\Delta_0(p)$, and the
amplitude $\Delta_\varphi(p)$, are plotted as a function of $p$.
One observes that $\Delta_\varphi(p)$ merges smoothly with
$\Delta_p(p)$ on the overdoped side. Consequently, $\Delta_0(p)$
is a smooth convex function of $p$ ranging from $\Delta_p$ to
$2\Delta_p$. Extrapolating $E_{dip}(p)$, it varies from $\sim
\Delta_p$ to $\sim 3 \Delta_p$ in the same range of $p$. At
optimal doping, $p=p_0$, the values are $\Delta_\varphi \simeq
\Delta_p/2$ and $\Delta_0 \simeq 3\Delta_p/2$. The dip position is
at $E_{dip}\simeq 2\Delta_p$, in agreement with ref. \cite{zaza1}.

Simple trends are found by plotting the parameters as ratios with
respect to $\Delta_p$, Fig.\,4b. We see that
$\Delta_\varphi$/$\Delta_p$, $\Delta_0$/$\Delta_p$ and
$E_{dip}$/$\Delta_p$ are increasing linearly as a function of $p$.
Moreover, if we write $p$ as the {\it excess} carrier density from
the minimum value of .05, i.e. the doping at the SC onset, we
obtain\,:
\begin{eqnarray}
  \frac{\Delta_\varphi}{\Delta_p} &\simeq& \frac{p}{2\,p_0} \\
  \frac{\Delta_0}{\Delta_p} &\simeq& 1+\frac{p}{2\,p_0} \\
  \frac{E_{dip}}{\Delta_p} &\simeq& 1.1 + \frac{0.9\,p}{p_0}
\end{eqnarray}
where $p_0 \simeq .11$ is the optimal doping. The first two are
obtained from the Taylor expansion of $\Delta_\varphi/\Delta_p$
while the third is from Eq.\,(13), using the value $\lambda =
1.06$. Thus $E_{dip}/\Delta_p$ shows nearly twice the slope as
$\Delta_\varphi/\Delta_p$, and varies from about $\sim 1$ to $\sim
3$ in the complete doping range.

Such a straightforward relationship between the parameters was not
expected. If $\Delta_\varphi$ is indeed the `condensation
amplitude', it continuously increases relative to the precursor
gap, $\Delta_p$, throughout the phase diagram. Since the latter
decreases linearly with doping, the order parameter is parabolic
shaped\,:
\begin{equation}
\Delta_\varphi = \frac{p\,\Delta_p(p)}{2\,p_0} \simeq 2.3\,
k_B\,T_c(p)
\end{equation}
With the previous hypotheses, Eq.\,(17) expresses a new precise
relation between the QP peak positions and the order parameter of
the SC transition.

To conclude the discussion on the results from the fits, it is
remarkable that the gap function, $\Delta(E)$, as displayed in
Fig.\,3, scales perfectly as a function of $p$ through the
variation of its amplitude, $\Delta_\varphi$, and the resonance
energy, $\lambda \Delta_0$. As with refs. \cite{miyakawa, zaza1},
we find no abrupt change in the QP spectrum, nor in its underlying
parameters, while spanning the carrier concentration. We conclude
that a critical point, if there is one, is situated at the right
end of the $T_c$ dome.

\vskip 2 mm

{\it Shape of the quasiparticle DOS}

\vskip 1 mm

The gap function $\Delta(E)$ is now uniquely determined for the
range of carrier concentration of interest ($.1<p<.24$). It is then
possible to fit the QP DOS with essentially one free parameter
($p$), aside from the broadening and the background slope. The
spectrum of Fig.\,2b, from Pan et al., was fitted by adjusting the
value of $p$, the final value being $p=0.175$ (see Fig.\,5). This
could be an approximate measure of the local value of the doping at
the surface of the sample, but the background slope adds some
uncertainty. Our objective here was mainly to reduce, as far as
possible, the number of free parameters.

\vskip 2 mm

\begin{figure}[h]
\centering
    \vbox to 4.8 cm{
    \epsfxsize=8.8 cm
   \epsfbox{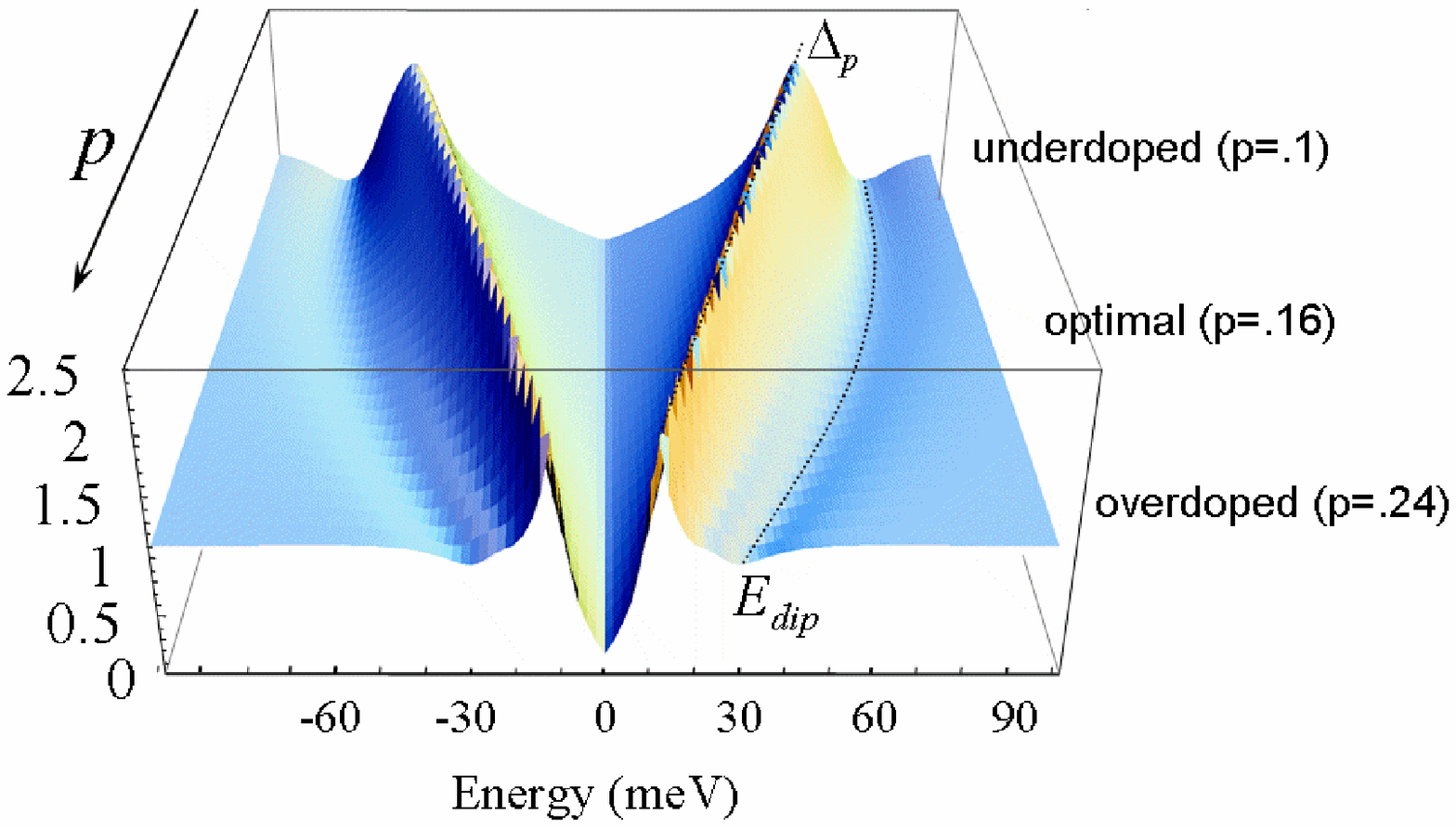}
    }
    \end{figure}

{\small Fig.\,5. Plot of the quasiparticle DOS evolution as a
function of $p$, the carrier density. The values of the parameters
used in the gap function are fixed from Fig.\,4a\,: the QP peaks
($\Delta_p$) follow a linear trend, while the dip position,
$E_{dip}$, lies along the curve given by (16). The difference
between the dip and the peak positions is $\sim
2\,\Delta_\varphi(p)$. Note the detailed DOS shape, varying
significantly from underdoped to overdoped sides.}

\vskip 2 mm

The variation of the QP DOS shape as a function of $p$ is shown as
a surface plot in Fig.\,5. As expected, the peak position follows
a linear law, while the dip follows the curve given by
$$E_{dip}\simeq 1.1 \Delta_p(p)+ 1.8 \Delta_\varphi(p)$$
as a direct consequence of the gap function (10). The detailed
shape of the QP DOS is not the same on each side of optimal doping
due to the relative position of $\Delta_0$ compared to $\Delta_p$.
Indeed, on the extreme underdoped side, where $\Delta_p
> \Delta_\varphi(p)$, the outer slope of the QP peaks is more
rounded, with a slight negative concavity between peak and dip. On
the overdoped side, $\Delta_p\simeq\Delta_\varphi(p)$, the DOS has
sharper peaks and dips, but also a positive concavity between
them. A closer look at the extreme overdoped case reveals a new
singularity, near $\Delta_0$ (a small kink). The kink is due to a
new extremum of $d\Delta(E)^2/dE$, but this prediction remains to
be verified experimentally.

The question as to whether we obtain `BCS behavior' on the
overdoped side can be addressed. In one sense we do\,: this limit
gives $\Delta_p(p)\simeq\Delta_\varphi(p)\simeq 2.3 \,k_B T_c$ and
a single parameter is then tied to $T_c$. However, the total
spectral gap $\Delta_0$ becomes twice too large ($\sim 2
\Delta_p$) so that the dip still persists, near $\sim 3 \Delta_p$,
which contradicts the BCS $d$-wave case. Thus, the precursor gap
maintains its influence when $p\rightarrow 2p_0$. In fact, the
results of Figs.\,4 and 5 clearly show that the gap function is
non BCS-like throughout the whole phase diagram.

Our model allows an order of magnitude for the value of $T^*$,
defined by the vanishing of the precursor gap. Using the data of
\cite{Campuzano, ding2, kam}, we find roughly that $2.8\, k_B\, T^*
\simeq \Delta_p$. Combined with Eq.\,14, we get\,:
$$ T^* \simeq 1.6\, (p_0/p)\, T_c$$
assumed valid for $T^*\geq T_c$. It follows that the two
temperatures merge at $p = 1.6\, p_0 \simeq .18$, perhaps remaining
merged for $p > 1.6\, p_0$, but this is an open question. Note that
the excess doping level at the ($T^*, T_c$) crossing point is .18,
while the absolute level is .23, i.e. well into the overdoped region
of the phase diagram.

\vskip 2 mm

{\it Model of the pseudogap state}

\vskip 1 mm

We propose a phenomenological description of how the gap function
should evolve when SC coherence is lost, i.e. the pseudogap state.
Since the model has fixed parameters for the SC state, we are
allowed one additional parameter to `force' the transition to the
PG state. There is very little data on the QP spectral change, and
we try only to be qualitative; no precise fit to a PG-type
spectrum is done.

\vskip 0 mm

\begin{figure}[h]
\centering
    \vbox to 3.8 cm{
    \epsfxsize=6.2 cm
  \epsfbox{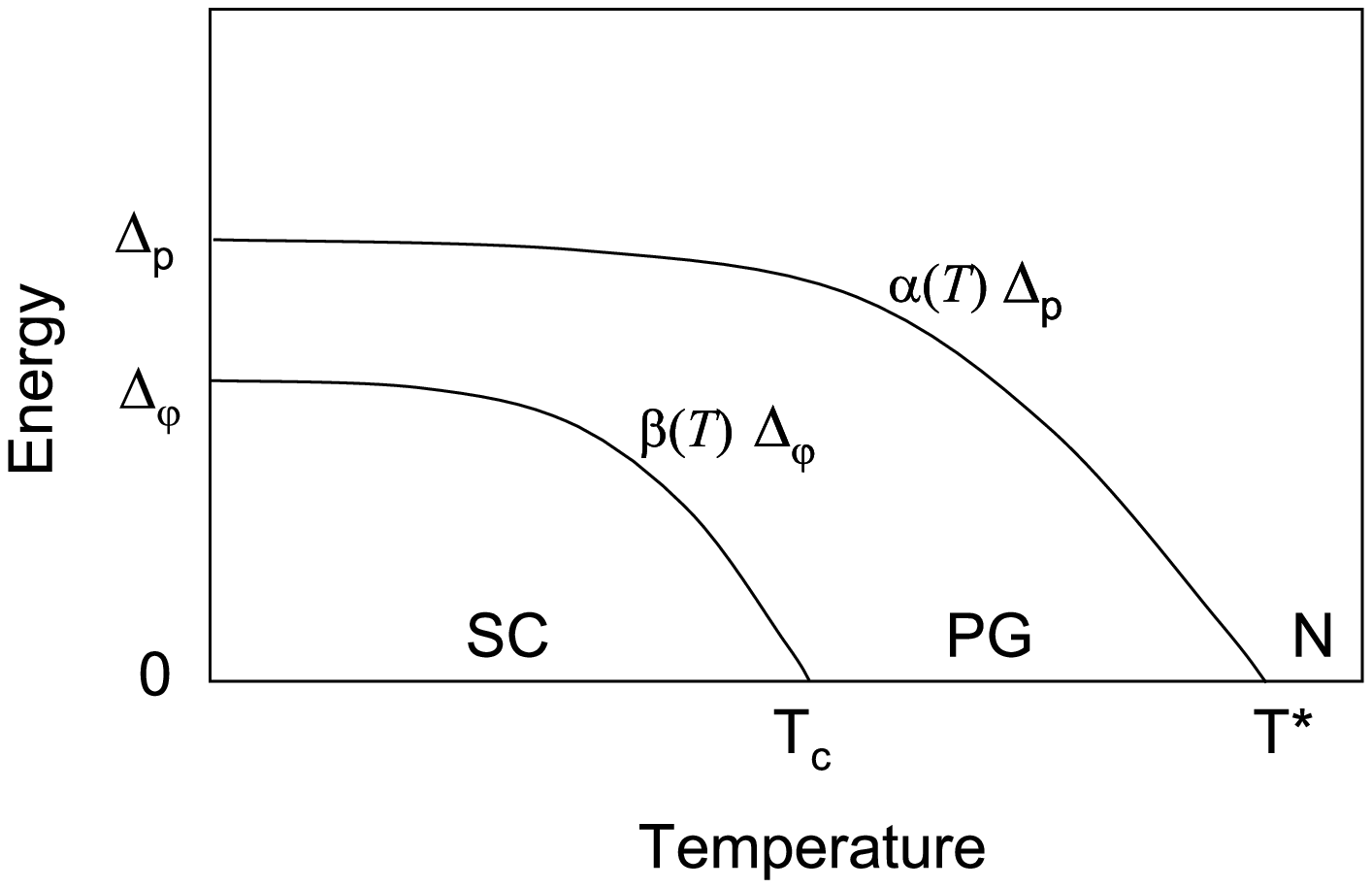}
    }
    \end{figure}

{\small Fig.\,6. Extension of the model to finite temperature
showing the $T$-dependence of $\alpha$ and $\beta$.}

\vskip 2 mm

Recall that the $T$-dependant BCS gap can be written $\Delta(T) =
\beta(T)\,\Delta(0)$, where $\beta(T)$ is a solution to the
finite-temperature gap equation and which verifies $\beta(0)$ = 1
and $\beta(T_c)$=0. A numerically exact, or simple interpolation
formula, can be found for $\beta(T)$. In an analogous way, we
write\,:
\begin{equation}
\Delta(E) = \alpha\ \Delta_{PG}\,(1-\beta) + \beta\ \Delta_{SC}(E)
\end{equation}
where $\Delta_{SC}(E)$ is the previous gap function and $\beta$,
$\alpha$ vary smoothly (but arbitrarily) from 1 to 0 as a function
of the strength of the perturbation (temperature, magnetic field,
etc.). We assume $\alpha$ and $\beta$ vanish at $T^*$ and $T_c$,
respectively, see Fig.\,6. The three region phase diagram is a
consequence of these two parameters, while $\Delta_{PG}$ and
$\Delta_{SC}$ are the characteristic zero-temperature energies. In
such a way, the three states are evidently\,:
\begin{eqnarray}
  \beta \neq 0 \ \ \ \ \alpha \neq 0 &\null& \ {\rm SC\ state}\nonumber \\
  \beta = 0 \ \ \ \ \alpha \neq 0 &\null& \ {\rm PG\ state}\nonumber \\
  \beta = 0 \ \ \ \ \alpha = 0 &\null& \ {\rm Normal\ state}\nonumber
\end{eqnarray}
In the case where $\Delta_{PG}$ is strictly identical to
$\Delta_p$ of the previous section, we get the form\,:
\begin{equation}
\Delta(E) = \alpha\ \Delta_p + \beta\ \Delta_\varphi (1 - g(E))
\end{equation}
so that the `condensation amplitude' is $\beta\,\Delta_\varphi$
and the `total' spectral gap is $\Delta_0 = \alpha\,\Delta_p +
\beta\, \Delta_\varphi$.

The QP-DOS shape corresponding to the PG state is an unsolved
problem, but there are many models \cite{varma, chakravarty,
levin2, emery, abanovSF, loktev, huscroft, ghosal, kwon,
atkinson2, pn, ohkawa, millis2, sachdev, abrikosov2,doucot}. There
is a strong indication of a broad-band self-energy in this case
\cite{norman,kam,chubukov,norman2,atkinson,huscroft}. Furthermore,
as discussed in the introduction, the PG-DOS is characterized by
the absence of quasiparticle peaks. Atkinson et al.
\cite{atkinson, atkinson2} have extended the earlier work of
Huscroft \& Scalettar \cite{huscroft} that a non-condensed Bose
state produces a peak-less pseudogap in the DOS. This has recently
been applied to the case of spatial disorder \cite{atkinson2}.
Franz \& Millis \cite{millis}, in an altogether different
approach, have found a pseudogap due to spontaneous currents, also
typical of a fluctuating pair system.

Our exceedingly simple model is to consider $\Delta_{PG}$ with a
finite imaginary part\,:
\begin{equation}
\Delta_{PG}=\Delta_p\,(1-i \delta')
\end{equation}
and to use Eq.\,(18) for the gap function. With $\beta$ lowered
from 1 to 0, the SC quasiparticle DOS evolves smoothly into the PG
shape (see Fig.\,7a). This type of DOS modification is seen in the
STS experiment as the tip is scanned within the vortex core
\cite{rennerB,phd,matsuba}. It is also seen as the effect of weak
disorder \cite{prlmaps}. Here $\delta'=.2$, which has a small, but
non-negligible, influence on the energy changes previously
discussed. Notice in Fig.\,7a that the dip features disappear as
$\beta\rightarrow0$, along with the QP peaks.

\vskip 2 mm

\begin{figure}[h]
\centering
    \vbox to 7.2 cm{
    \epsfxsize=6.4cm
  \epsfbox{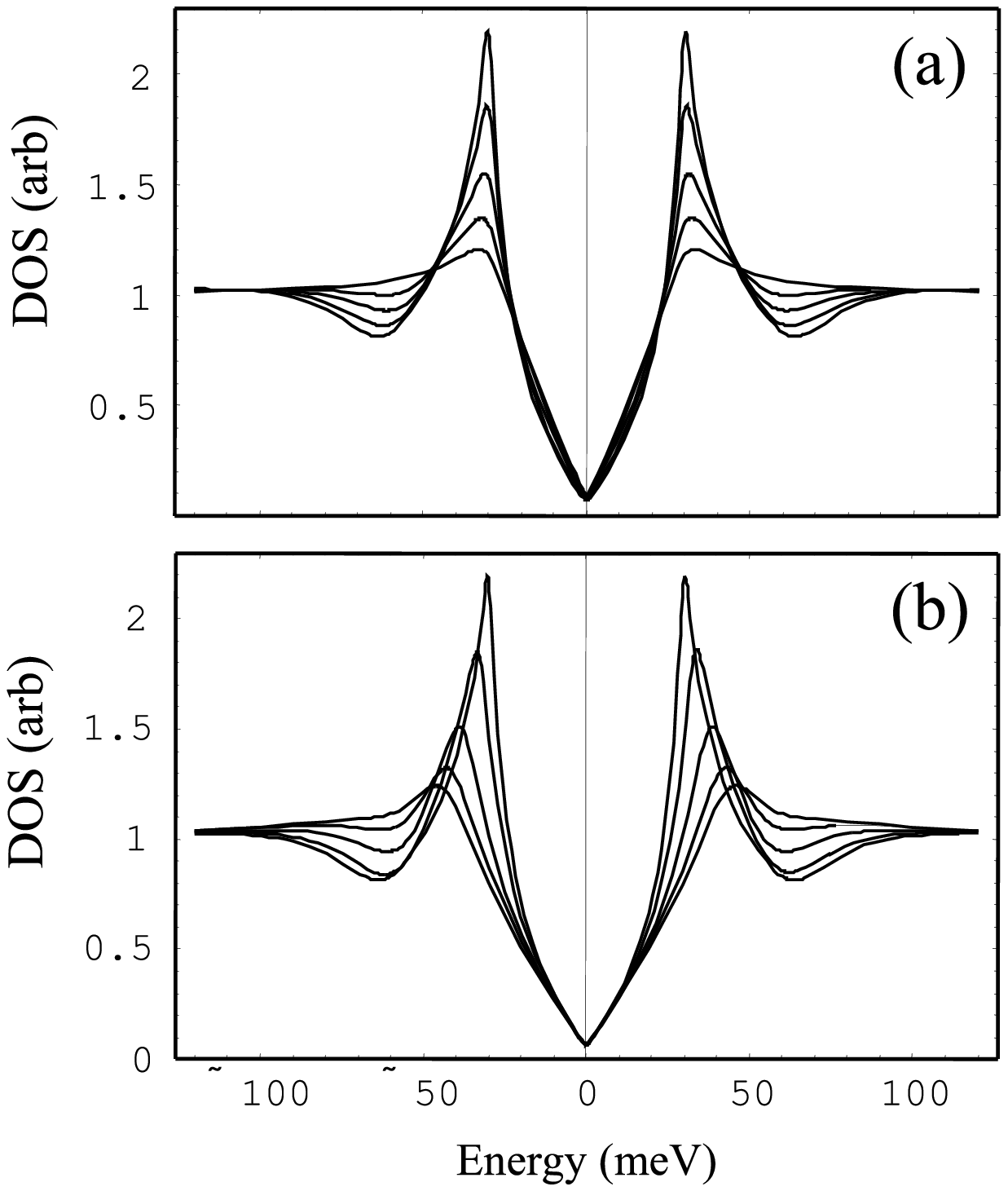}
    }
    \end{figure}

{\small Fig.\,7. Variation of the DOS with $\beta$ varying from 1
(SC state) to 0 (PG state). The parameters are $p$=.18, giving
$\Delta_p = 30$ meV and $\Delta_\varphi$ = 17.8 meV, in Eq.\,18,
and $\delta'=.2$, in Eq.\,20. The broadening parameters for the SC
state are\,: $\delta=.06$, $\Gamma=.03\,\Delta_p$. In (a) ${\rm
Re}\, \Delta_{PG} = \Delta_p$, in (b) ${\rm Re}\,\Delta_{PG}=
40$\,meV, i.e. larger than $\Delta_p$.

These variations simulate the loss of SC coherence in the case (a)
in the vortex core \cite{rennerB, phd} and (b) due to strong
disorder \cite{nanomaps, howald, pan}.}

\vskip 4 mm

In Fig.\,7b we let $\beta$ vanish but here ${\rm Re}\,
\Delta_{PG}>\Delta_p$\,: the pseudogap is slightly larger than the
quasiparticle peak energy (40 meV as compared to 30 meV). Now the QP
peaks move progressively out to a higher energy, as expected from
Eq.\,(18), while being attenuated. This particular evolution is
strikingly similar to STS observations
\cite{nanomaps,howald,pan,sugimoto}, and is generally attributed to
disorder (inhomogeneity). In our model, the {\it decrease} of doping
level on the local scale could give this effect, the result being a
wider pseudogap. It corresponds to the progressive lowering of the
$p$-value in the phase diagram of Fig.\,4a. However, a disorder
potential is expected to directly affect the SC amplitude, hence
${\rm Re}\,\Delta_{PG}$ could vary in different regions of the
surface \cite{nanomaps}. Consequently, the SC state to PG state
transition energy depends on the local change of the precursor gap,
(${\rm Re}\,\Delta_{PG}-\Delta_p$), and on $\Delta_\varphi$. An
estimate of this energy is given in the following section.

\begin{figure}[h]
\centering
    \vbox to 6.1 cm{
    \epsfxsize=8cm
 \epsfbox{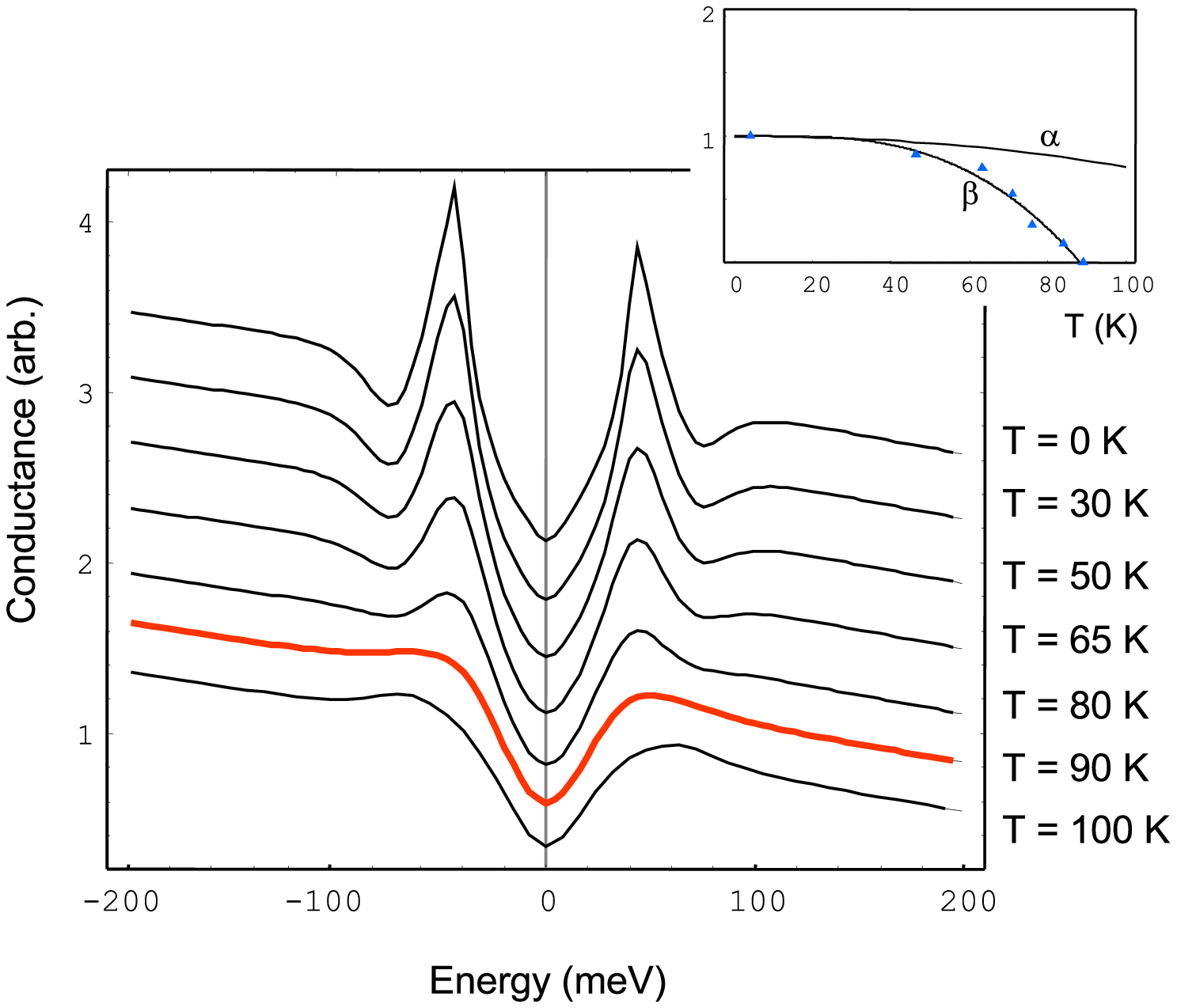}
    }
    \end{figure}

{\small Fig.\,8. Simulation of the tunneling conductance with
temperature. The value of $\beta$, varying from 1 (SC state) to 0
(PG state), was inferred from the data \cite{rennerT}, shown in
the upper inset. The parameters are $p$=.14, giving $\Delta_p =
42$ and $\Delta_\varphi$ = 17.6 meV, in Eq.\,18; all others are
the same as in Fig.7a. Here we assume ${\rm Re}\,\Delta_{PG} =
\Delta_p$, but $\alpha$ in Eq.\,18 decreases slightly from unity
(see inset). The highlighted spectrum corresponds to $T_c$.}

\vskip 2 mm

The well-known experiment of Renner et al. \cite{rennerT} gives
the quasiparticle conductance spectrum as a function of rising
temperature. In Fig.\,8 we plot a simulation of the temperature
dependance of the QP spectra using Eqs.\,(4,\,9) and the gap
function (19). The standard broadening of $N_S(E)$ due to Fermi
statistics is taken into account \cite{bergeal}. The value of
$\beta(T)$ is inferred from the data (upper inset, Fig.\,8) and it
vanishes at $T_c\simeq89\,$K. The $p$ value is chosen such that
$\Delta_p(p)\simeq 42\,$ meV. As is well established from many
experiments, the PG persists well above $T_c$, up to the
temperature $T^*$ previously discussed. Our model, as shown in
Fig.\,8, gives a simple qualitative description of the QP DOS with
rising temperature.


\section{III. Interpretation of the energies}


We turn to the question of interpreting the two energy scales
($\Delta_p$, $\Delta_\varphi$) that are the main ingredients of
the gap function $\Delta(E)$, as expressed in (19). The foregoing
work shows that, for large QP energies and along the anti-nodal
direction, the total spectral gap is $\Delta(E) \simeq \Delta_0 =
\Delta_p + \Delta_\varphi$, where $\Delta_p$ defines the QP peak
positions, and the new parameter, $\Delta_\varphi$, is
proportional to $T_c(p)$. We now consider the energy changes
involved in both the normal to SC and the PG to SC transitions. In
the latter case we use (20) for the PG gap, with the assumption\,:
Re $\Delta_{PG}= \Delta_p$, with a small imaginary part.

Consider first the BCS case of a constant isotropic gap of value
$\Delta$ in the SC spectrum. The approximate change in energy is
given by the familiar integral\,:
\begin{eqnarray}
  \left(\rm E_N - E_{SC}\right)_{BCS} &=& N_n(0)\,\int_{0}^{2\pi} \frac{d\theta}{2\pi} \int\,
   d\epsilon_k\,\left(\frac{\epsilon_k^2}{E_{\bf k}} - |\epsilon_k|
   \right)\nonumber \\
  &+& N_n(0)\,\int_{0}^{2\pi} \frac{d\theta}{2\pi} \int_{}^{}\,d\epsilon_k\,\frac{\Delta}{2\,E_{\bf k}}
\end{eqnarray}
The first and second terms \cite{degennes} are the change in kinetic
and potential energies, respectively, while the variation of the
chemical potential is neglected. The high energy cut-off is
unnecessary since the integrand vanishes faster than $1/\epsilon_k$.
The standard result is\,:
$$\left(\rm E_N - E_{SC}\right)_{BCS} = \frac{1}{2}\
N_n(0)\,\Delta^2$$ or equivalently, putting $N_p = N_n(0)\Delta/2$
as the number of pairs,
\begin{equation}
   \frac{\left(\rm E_N - E_{SC}\right)_{BCS}}{N_p} = \Delta
\end{equation}
The conventional interpretation is therefore straightforward\,: the
gap in the QP spectrum is equivalent to the energy gain per pair in
the N $\rightarrow$ SC transition. The latter intensive quantity is
related to the mean-field critical temperature, $\Delta = 1.76\,
k_B\, T^{MF}_c$.

In the unconventional case, we must go beyond the mean-field due
to three possible effects\, (a) weak phase-stiffness and 2D
electronic structure (Kosterlitz-Thouless limit), (b) proximity to
Bose-Einstein condensation and (c) coupling to a collective mode
(additional degrees of freedom). Denoting $E_\varphi$ the energy
gain due to one of these processes\,:
$$\left(\rm E_N - E_{SC}\right) =
\left(\rm E_N - E_{SC}\right)'_{BCS} + \,E_\varphi$$
where $(\rm E_N - E_{SC})'_{BCS}$ is given by (21), but using
the complete gap function $\Delta(E_k)$ (i.e. an extended BCS
energy). The latter is calculated by using (18), with $\alpha =
\beta = 1$, and integrating (21)\,:
$$\left(\rm E_N - E_{SC}\right)'_{BCS} \simeq \frac{1}{2} \times
\frac{1}{2} N_n(0) \Delta_p^2 $$ where the extra factor of 1/2 is
from the angular integral for the $d$-wave case.
Here one would have expected a {\it higher gap value}
than $\Delta_p$ (between $\Delta_p$ and
$\Delta_0$), but the anti-resonant shape of $\Delta(E)$ leads to
this simple result and is discussed below. The total energy change
from the normal to SC states is then\,:
\begin{equation}
\left(\rm E_N - E_{SC}\right) = \frac{1}{4} N_n(0) \Delta_p^2 +
\,E_\varphi
\end{equation}

If one estimates the new number of gapped pairs as $N_p =
N_n(0)\Delta_p/2\sqrt{2}$, dividing through by $N_p$ gives\,:
\begin{equation}
   \frac{\left(\rm E_N - E_{SC}\right)}{N_p} = \frac{1}{\sqrt{2}}
   \Delta_p + \frac{E_\varphi}{N_p}
\end{equation}
This equation matches our expression\,: $\Delta_0 = \Delta_p +
\Delta_\varphi$, with the identification\,:
$$\frac{1}{\sqrt{2}} \Delta_0 = \frac{\left(\rm E_N -
E_{SC}\right)}{N_p} {\rm \ \ \ and \ \ \ } \frac{1}{\sqrt{2}}
\Delta_\varphi = \frac{E_\varphi}{N_p} \ \ \ .$$ Neglecting the
small imaginary part of $\Delta_{PG}$, we find the energy change
(N$\rightarrow$PG)\,:
\begin{equation}
\frac{\left(\rm E_N - E_{PG}\right)}{N_p} = \frac{1}{\sqrt{2}}
   \Delta_p
\end{equation}
and finally, combining (23) and (24),
\begin{equation}
\frac{\left(\rm E_{PG} - E_{SC}\right)}{N_p} = \frac{E_\varphi}{N_p}
\end{equation}
Thus the $E_\varphi$ term is precisely the energy gain of the SC
state with respect to the PG state\,: the condensation energy. The
intensive quantity, $E_\varphi/N_p$, is of the order of $k_B\,T_c$.
Since $E_\varphi$ is not a `pairing' energy, we put $N_p = N/2$ and
obtain\,:
$$\Delta_\varphi = 2\sqrt{2}\ \frac{E_\varphi}{N} \simeq
2\sqrt{2}\ k_B\,T_c \simeq 2.8 \ k_B\,T_c$$ in qualitative agreement
with the fits. We have thus shown that the heuristic parameter
$\Delta_\varphi$ is consistent with a condensation energy,
$E_\varphi$, going beyond the BCS mean-field value.

These arguments cannot prove whether $\Delta_p$ is a pairing gap
(pre-formed pairs above $T_c$) or a competing order gap (spin
density wave, etc.). It is compatible with many models where states,
in the energy range $\sim\Delta_p$, are first removed at the Fermi
level. However, our model suggests that the new quantity,
$\Delta_\varphi = 2\sqrt{2}\,E_\varphi/N$, is not the opening of a
second gap at the SC transition. Indeed, in a first approximation,
the total change in KE \& PE (i.e. given by the BCS Eq.\,(21)) is
identical with the pseudogap contribution\,:
\begin{equation}
\left(\rm E_N - E_{SC}\right)'_{BCS} \simeq \left(\rm E_N -
E_{PG}\right)
\end{equation}
with the consequence\,:
\begin{equation}
\left(\rm E_{PG} - E_{SC}\right)'_{BCS} \simeq 0
\end{equation}
However, as previously noted, this is compatible with experiment\,:
using STS, the PG width measured at the vortex core is about the
same magnitude as the SC gap between the vortices. The condensation
energy term, $E_\varphi = {\rm (E_{PG} - E_{SC})}$, is therefore
apart, and must be due to additional degrees of freedom beyond the
mean field.

Behind these results (26-28) is a non-trivial mouvement of states
in the PG $\rightarrow$ SC transition \cite{normanke}. Consider
again Fig.\,3 plotting the shape of the gap function, and in
particular, the role of $\Delta_\varphi$. The states at higher
energy, $E \gg E_{dip}$, are removed ($\Delta(E) \sim \Delta_p +
\Delta_\varphi$) while they are added near the anti-resonance
($\Delta(E_0) \sim \Delta_p - \Delta_\varphi$). Due to a detailed
balance of energy states in the integrand $f(\epsilon_k)$ of (21),
the effect cancels out and we are left with the result Eq.\,(27).
This is explicit in Fig.\,9 where $f(\epsilon_k)$ is plotted as a
function of $\epsilon_k$ (along $\theta$ = 0). We use the previous
gap expression with $\alpha=1$\,:
$$\Delta(E_k) = \Delta_p + \beta\, \Delta_\varphi\, (1-g(E_k))$$ and
we compare directly the cases of $\beta$=1, and $\beta$=0. As
expected, the latter curves cross in such a way as the subtended
areas are the same. In our view, this removal of states at higher
energy, also seen in the spectral function (Sec. IV), is a
significant aspect of the problem. A microscopic theory should
arrive at a clear dependence of $E_\varphi$ on this effect.

\vskip .5 mm

\begin{figure}[h]
\centering
    \vbox to 3.8 cm{
    \epsfxsize=6.2cm
  \epsfbox{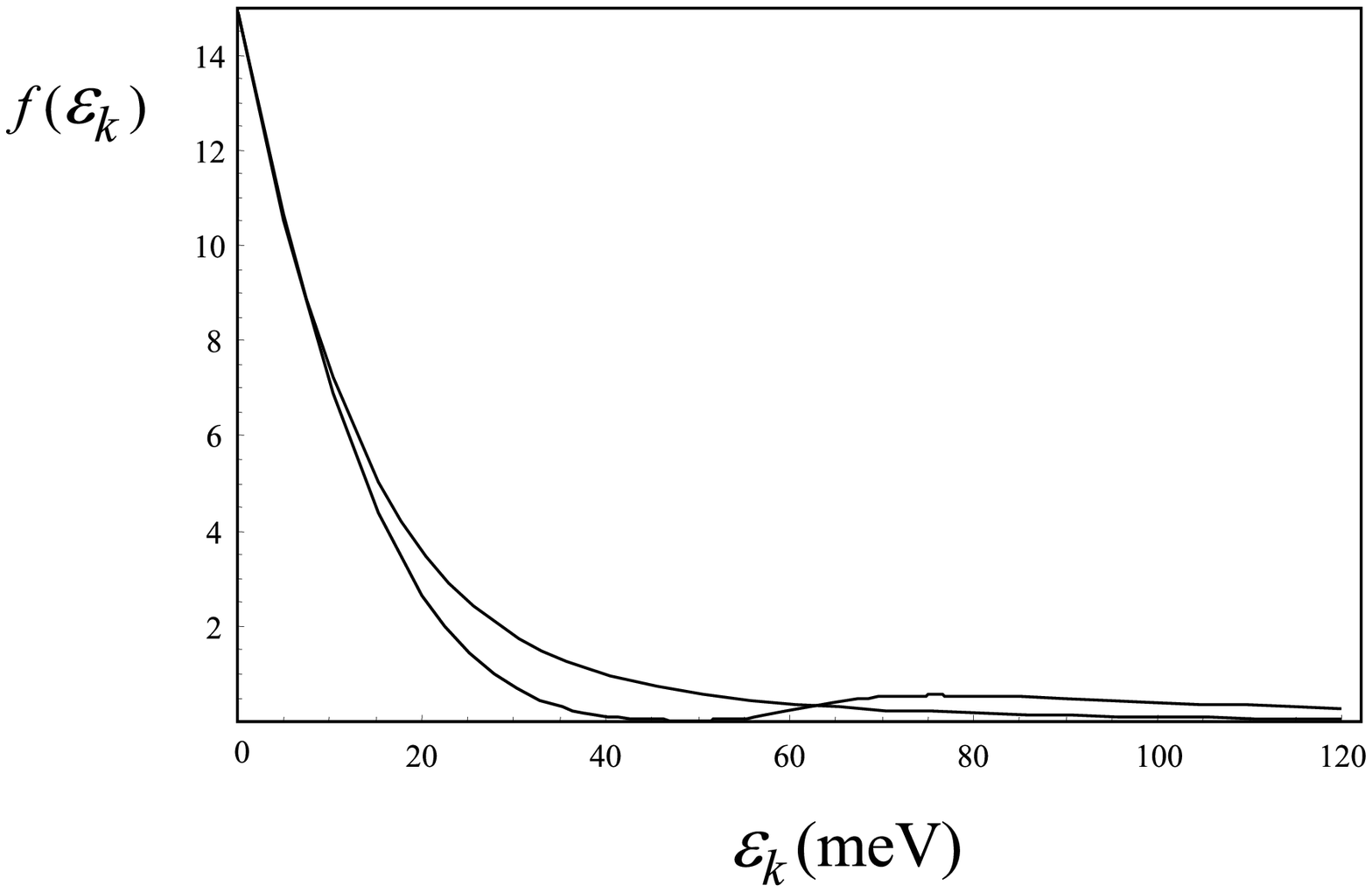}
    }
    \end{figure}
{\small Fig.\,9. Plot of the integrand, $f(\epsilon_k)$, of the
BCS energy equation (21) in units of $N_n(0)/(2\,\pi)$ and for
$\theta=0$. The two curves compare the case of a constant gap,
with $\Delta_p = 30$\,meV ($p=.18$), with the case of the gap
function, from Fig.3. The latter curve shows a pronounced dip -
hump feature, but the subtended areas are the same.}

\vskip .3 cm

It is possible that, in some instances, the pseudogap is larger
than the QP peak position\,: $\Delta_{PG}>\Delta_p$. The total
energy change, Eq.\,(23), remains valid but (26) becomes\,:
$$
\left(\rm E_{PG} - E_{SC}\right) = - \frac{1}{4} N_n(0)
(\Delta_{PG}^2-\Delta_p^2) + \,E_\varphi
$$
The change in gap value now opposes the previous condensation energy
term. (In fact, if $\Delta_{PG}$ is large enough, then no transition
to the SC state occurs\,: the above vanishes identically.) A
disorder potential could be such a cause, when amplitude
fluctuations are expected. Then, the above equation implies a change
in the number of gapped states at the PG $\rightarrow$ SC transition
and, in this case, $E_\varphi$ need not strictly be the order
parameter.

The question for further work is to interpret the condensation
amplitude, $\Delta_\varphi = 2\sqrt{2}\,E_\varphi/N$. A likely
candidate is the phase stiffness $J$ relevant to the
Kosterlitz-Thouless (KT) transition\,:
$$E_\varphi/N = (\pi/2) J =  \ k_B\,T^{KT}_c$$ For free-electrons
in two dimensions, one shows that \cite{pn}\,: $J = n \hbar^2/4m$,
the main point being that $J \propto n$, the particle density.
From the previous fits to the data, we inferred that
$\Delta_\varphi = (p/2p_0)\,\Delta_p$, so that it is at least
proportional to the excess density $p$. If the KT mechanism is
involved, then the principal quantity of the SC gap function is
predicted to be\,:
$$\Delta_\varphi = \sqrt{2}\,\pi\,\,J = 2\sqrt{2}\, \ k_B\,T^{KT}_c $$
with a new condition on the  phase-stiffness\,:
$$\sqrt{2}\,\pi\,J = \frac{p\,\Delta_p}{2\,p_0}$$
The latter result is satisfactory; the phase rigidity is now
proportional to both the carrier density and the precursor
interaction. Put another way, using $p' \propto N_n(0)\,\Delta_p$
as the density of gapped states, we have the simple constraint\,:
$J \propto p\,p'$. Since the latter is valid for all $p$ (ranging
from 0 to 2$\,p_0$), the dome-shaped SC phase diagram is obtained,
i.e. $p\,p' \propto T_c(p)$.

\vskip .5 mm

In summary, without committing to a specific camp in the
high-$T_c$ problem, the total spectral gap is $\Delta_0 = \Delta_p
+ \Delta_\varphi$, where $\Delta_p$ is the precursor gap and and
the amplitude $\Delta_\varphi$ is proportional to the condensation
energy. The latter term, and its link to the phase diagram, points
however to the mechanism of `pre-formed' pairs followed by KT
condensation. This situation of two energy scales was suggested in
the work of G.\,Deutscher \cite{deutscher}, where the QP tunneling
gap ($\Delta_p$) is distinct from the coherence gap
($\Delta_\varphi$ ?) as inferred by the Andreev effect.

\vskip 2 mm

{\it Further comments on the gap function}

\vskip .5 mm

Consider the question of why the particular QP spectrum, $E_{\bf
k}$, with the anti-resonant gap function $\Delta(E_{\bf k})$, is
needed to produce the correct QP-DOS. First, the resonance is
shown to cause a shift in the quasiparticle states, at the PG to
SC transition, from higher energy to lower energy, manifested by
the sharp QP peaks and dip features. Nevertheless, a simple answer
for the energy change is obtained. All these points indicate that
the condensate itself is non-trivial \cite{normanke}. Furthermore,
the new quantity $\Delta_\varphi$ enters the resonance position
($E_0 \simeq \Delta_0$), its width and its amplitude (2$
\Delta_\varphi$); a property shown to be empirically correct
throughout a wide doping range. Then it seems unlikely that the
resonance is due to the coupling to a quasi-independent collective
mode. In the latter case, the width of the resonance would reflect
the collective mode damping, while its amplitude, the strength of
the coupling. Here these parameters are related to the coherence
amplitude, $\Delta_\varphi \propto E_\varphi/N$, possibly tied to
the phase-stiffness $J$, and hence to the condensate itself.

The reason for the energy dependence of the gap function is so far
unknown from first principles. Aside from strong-coupling theory
\cite{schrieffer}, an $E$ dependent gap appears in McMillan's
proximity effect \cite{mcmillan}, two-band superconductivity
\cite{suhl}, and the asymmetric particle-hole case\,\cite{hirsch}.
However, a highly relevant example is the detailed work of
\cite{nozieresBE} on the cross-over from BCS to Bose-Einstein
condensation; but the gap has no resonance effect there.

The detailed form of the gap function, or self-energy as in
Section IV, represents a novel pairing interaction. It is possible
that there is a strong feedback effect between the precursor
state, with gap $\Delta_{PG}$, and the freezing of phase
fluctuations, of amplitude $\Delta_\varphi$. Then there is a
complete renormalization of the gap function $\Delta(E_k)$, where
$E_k$ is the `exact' quasiparticle spectrum (final state). The
effect is strong enough so that if $E_k \simeq \Delta_0$, the
pairing energy is reduced, $\Delta \simeq \Delta_p -
\Delta_\varphi$, compared to the low-energy value $\Delta \simeq
\Delta_p$. On the contrary, at higher energy, the interaction is
strong $\Delta \simeq \Delta_p + \Delta_\varphi$. This type of
variation is at the heart of the QP-DOS shape.



\section{IV. Spectral function}

A different, yet equivalent, view of the same problem is given by
the spectral function, $A({\bf k}, E)$. It is more fundamental
since there is no sum over momenta, hence an ideal quasiparticle
with wave-vector ${\bf k}$ is directly a Dirac peak at the
position $E=E_{\bf k}$. It is also the fundamental quantity
describing the ARPES measurement.

Here we discuss $A({\bf k}, E)$ as given by equation (8), with the
quasiparticle dispersion (7), the total gap function\,:
$$\Delta(E_{\bf k}) = \alpha \Delta_{PG}\,(1-\beta) + \beta\,
\Delta_{SC}(E_{\bf k})$$ and with $\Delta_{SC}(E_{\bf k})$ the
usual resonant form given in (10). The gap broadening parameters
are as in Sec. II, i.e. $\Delta_p \rightarrow
\Delta_p\,(1-i\delta)$ and $\Delta_{PG} \rightarrow
\Delta_p\,(1-i\delta')$ in $\Delta_{SC}$ and $\Delta_{PG}$,
respectively. The final broadening parameter is $\Gamma_{Dynes}$,
and all three have been determined by the fits in Sec.\,II.

Since $A({\bf k}, E)$ is calculated along the real $\epsilon_k$
axis, it follows that our dispersion law $E_{\bf k}=E_{\bf
k}(\epsilon_k)$, Eq. (7), is complex. Considering the anti-nodal
direction, the simple model gives\,:
\begin{equation}
A({\bf k}, E) = \frac{1}{\pi} \ \frac{Im\,E_{\bf k}}{(E-Re\,E_{\bf
k})^2 + Im\,E_{\bf k}^2}
\end{equation}
which has the expected Lorentzian form with a peak at $Re E_{\bf k}=
E$ and half-width $Im\,E_{\bf k}$. Such an expression neglects
many-body effects that are important at higher energy
\cite{norman2,norman3}.

\vskip 1 mm

\begin{figure}[h]
\centering
    \vbox to 10.6 cm{
    \epsfxsize=5.8 cm
    \epsfbox{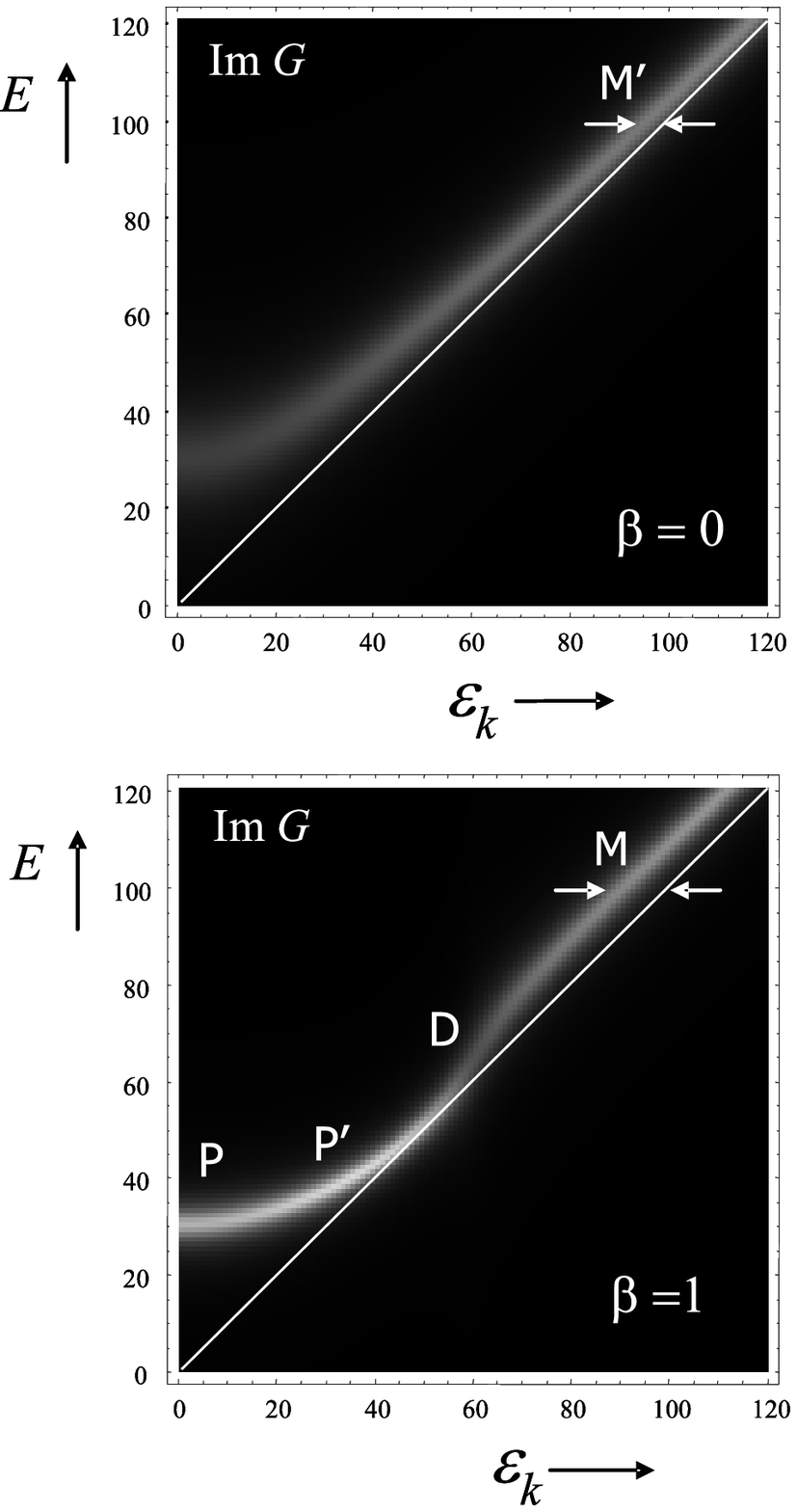}
    }
    \end{figure}

{\small Fig.\,10. Direct comparison of the spectral functions of
the PG state (upper panel) and of the SC state (lower panel),
viewed as density plots in the ($E,\epsilon_k$) plane. The gap
parameters correspond to $p=.18$\,: $\Delta_p=30$\,meV and
$\Delta_\varphi=17.8$\,meV. Other relevant parameters are\,:
$\delta=.06$, $\delta'=.2$ and $\Gamma=.06\,\Delta_p$.\\
The dispersion law, Re$\,E_k$, which is hyperbolic in the PG case,
has a distinctive kink (at D) in the SC case, causing the dip
feature in the DOS. States are removed at high energy from the PG
spectral density (compare M and M'), and are added at low energy
(between P and D), in the SC spectrum.}

\vskip .2 cm

The spectral function (29) is presented in grey scale in Fig.\,10
in the ($E,\epsilon_k$) plane and for the case where $p=.18$. Note
that the gap function is an explicit function of the QP energy,
$E_{\bf k}$, and not $\epsilon_k$; thus one must first find
numerically the (complex) dispersion law $E_{\bf k}(\epsilon_k)$
in order to apply (29). On the lower panel of Fig.\,10 we give the
spectral function for the SC state and compare directly with the
one for the PG state (upper panel). For the SC state, the visibly
sharp peak position describes a continuous curve that deviates
significantly from the BCS hyperbolic law. The hyperbolic case is
seen in the upper panel, since it is inherent in the model for the
pseudogap. For small $\epsilon$ both curves begin near the gap
value $E\sim\Delta_p=30\,$ meV, as expected. The QP peak position
(SC state) begins (at P) with a slightly smaller slope than the
BCS case and nearly reaches the $E = \epsilon_k$ diagonal. This is
followed by a sharp kink (near $\epsilon_k =$ 60 meV, at D) before
joining the expected asymptote for a gap value of $\Delta_0 \simeq
48\,$meV. All of these effects are totally absent in the BCS
framework.

The kink feature in the dispersion law corresponds to the dip in
the QP-DOS. One can read its position, D, either on the $E$ axis
or on the $\epsilon_k$ axis\,: in the former, $E_{dip}\sim
\Delta_p + 2 \Delta_\varphi \sim 64\,$meV, while in the latter
$\epsilon_k\sim 2 \sqrt{\Delta_0 \Delta_\varphi}\sim\,$ 60 meV
(taking the point of steepest slope). The anti-resonance in the
gap function is revealed here by the near touching of the diagonal
by the dispersion law at the energy $E \sim \Delta_0 \sim
48$\,meV. (At this point $\epsilon_k \sim E$.) Concerning the peak
intensity, the sharpest QP peaks are not at $\epsilon_k = 0, E =
\Delta_p$ (P in Fig.\,10) but at slightly higher energy, near
40\,meV (at P'). At the kink, the peak intensity drops
significantly (highest broadening), giving a quite different
picture of the peak to dip feature.

\begin{figure}[h]
\centering
    \vbox to 3.9 cm{
    \epsfxsize=6cm
  \epsfbox{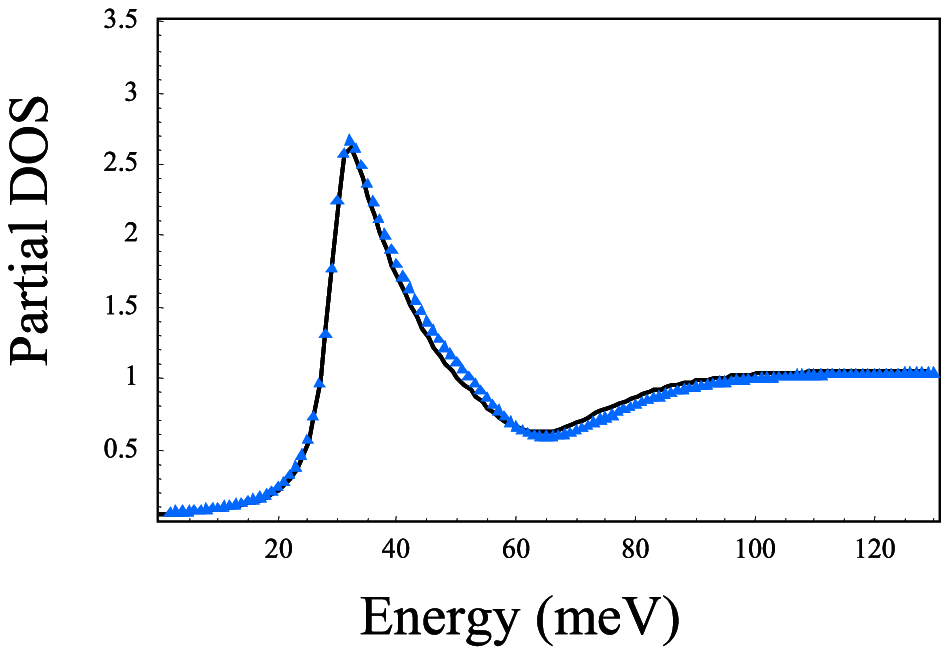}
    }
    \end{figure}

{\small Fig.\,11. Tunneling DOS (along $\theta=0$) calculated
using the exact formula (9), solid line, and the discrete sum over
$\epsilon_k$ of the spectral density of Fig.\,10, triangular
dots.}

\vskip .3 cm

Although ARPES can only access the occupied states, contrary to
tunneling spectroscopy, the two techniques probe the same spectral
function (Fig.\,10). Indeed, integrating $A({\bf k}, E)$ over
$\epsilon_k$, for a given $E$, yields the partial tunneling DOS,
$n_s(E,\theta)$, as given by Eq.\,(6). In ARPES one can fix ${\bf
k}$, hence $\epsilon_k$, and measure $A({\bf k}, E)$ as a function
of $E$ (energy distribution curve). Moreover, the dispersion of
the QP states with ${\bf k}_\parallel$ can be probed. Although
these two techniques have different spatial and energy
resolutions, it would be of high interest to establish the common
spectral function that matches both the ARPES and the tunneling
data.

As a simple check of the tunneling DOS, in Fig.\,11 we plot both
the exact partial DOS (for $\theta=0$), using eq.(9), and the
direct sum over states of the spectral density of Fig.\,9. Given
the crude integration method in the latter case, the agreement is
satisfactory. All experimentally observed features are found\,:
the pronounced QP peak and the dip, as well as the states within
the gap from the Dynes broadening.

The upper panel of Fig.\,10 shows the spectral function
corresponding to the PG state, or $\beta=0$ in the gap function,
for direct comparison. Notice the significant smearing near
$\epsilon_k \sim 0$, or $E\sim \Delta_p$\,: as expected, there are
no well-defined quasiparticle peaks. The overall dispersion shape
is still hyperbolic and follows $E_k \sim
\sqrt{\epsilon_k^2+\Delta_p^2}$ as imposed by the model. The
asymptotic line is closer to the diagonal as compared to the SC
case, where the total high-energy spectral gap is larger, i.e.
$E_k \sim \sqrt{\epsilon_k^2 + \Delta_0^2}$ (compare M' with M in
Fig.\,10). This gives a direct view of the mouvement of the states
in the PG $\rightarrow$ SC transition\,: states well above the PG
gap, for $E_k>E_{dip}$ (upper panel), are removed and a high
density of quasiparticle states exist, for $E_k<E_{dip}$, in the
SC state (lower panel).

\vskip 1.5 mm

{\it Quasiparticle self-energy}

\vskip .5 mm

The final objective of this work is to show that the $E$-dependant
gap function is equivalent to a self-energy. In the SC state, the
carriers of energy $\epsilon_k$ are modified by the presence of
$\Sigma_{\bf k}(\epsilon_k)$\,:
\begin{equation}
A({\bf k}, E) = \frac{1}{\pi} \ Im \frac{1}{E-\epsilon_k-\Sigma_{\bf
k}(\epsilon_k)}
\end{equation}
Thus, setting $\Sigma=0$ gives the normal state spectrum, while for
ideal BCS quasiparticles it is with $\Sigma_{\bf
k}(\epsilon_k)=\sqrt{\epsilon_k^2 + \Delta^2}-\epsilon_k$. Finally,
the Dynes lifetime broadening is obtained if $\Sigma$ has an
imaginary part (Im $\Sigma=\Gamma$).

For the general case, where the SC state is given by the gap
function, comparing (30) with our original expression (8), gives\,:
\begin{equation}
\Sigma_{\bf k}(\epsilon_k) = E_{\bf k}(\epsilon_k) - \epsilon_k
\end{equation}
However, the final state QP energy, $E_{\bf k}(\epsilon_k)$, is a
solution of the implicit equation $E_{\bf k} = \sqrt{\epsilon_k^2 +
\Delta_{\bf k}(E_{\bf k})^2} + i\,\Gamma$, and thus an analytical
expression for $\Sigma_{\bf k}(\epsilon_k)$ cannot be given. The
computation of $E_{\bf k}(\epsilon_k)$ is quite straightforward,
leading to the complex valued $\Sigma_{\bf k}(\epsilon_k)$ and hence
to the spectral function (30). Formally, the quasiparticle resonance
position is given by $Re E_{\bf k} = \epsilon_k + Re \Sigma_{\bf
k}(\epsilon_k)$, and its width by $Im E_{\bf k} = Im \Sigma_{\bf
k}(\epsilon_k)$. Note that using the definition of $E_{\bf k}$, and
the simple expression (31), yields\,:
\begin{equation}
\Sigma_{\bf k}(\epsilon_k) = i \Gamma + \frac{\Delta_{\bf k}(E_{\bf
k})^2}{E_{\bf k} - \epsilon_k + i \Gamma}
\end{equation}
which is similar to the form considered by \cite{hoogenboom,
norman2}. Thus, the $E$-dependent gap function and the self-energy
function are equivalent.

The self-energy is plotted in Fig.\,12a using the same parameters
as those of Figs.\,10 and 11. The darker line corresponds to the
SC case using the resonant gap function, the gray line is for the
PG model, with a constant gap ($\sim \Delta_p$). For small
$\epsilon_k$ both curves begin at the value\,: ${Re}\,\Sigma(0)
\simeq
 \Delta_p \simeq 30\,$ meV, as expected. However, the SC
curve descends sharply towards zero, as compared to the PG one,
and reaches a minimum near $\epsilon_k\simeq
2\sqrt{\Delta_p\,\Delta_\varphi}$. This corresponds to the
anti-resonance position in the gap function ($E = \Delta_0$). The
self-energy then rises abruptly, crosses the PG reference curve
and, with a further change in concavity, decays as a function
$\epsilon_k$. The dip position in the QP-DOS is found by the
maximum slope of ${\rm Re}\, \Sigma(\epsilon_k)$; it is near the
crossing point at $\epsilon_k\simeq
2\sqrt{\Delta_0\,\Delta_\varphi}$. Note that this dip position
coincides with the highest broadening, as revealed by ${\rm Im}\,
\Sigma(\epsilon_k)$ in the lower panel of Fig.\,12a. The lowest
broadening is at $\epsilon_k \sim 40\,$ meV, giving the highest QP
peaks in the spectral function, and the wide peaks seen in the
DOS.

\begin{figure}[h]
\centering
    \vbox to 4.4 cm{
    \epsfxsize=8.5cm
  \epsfbox{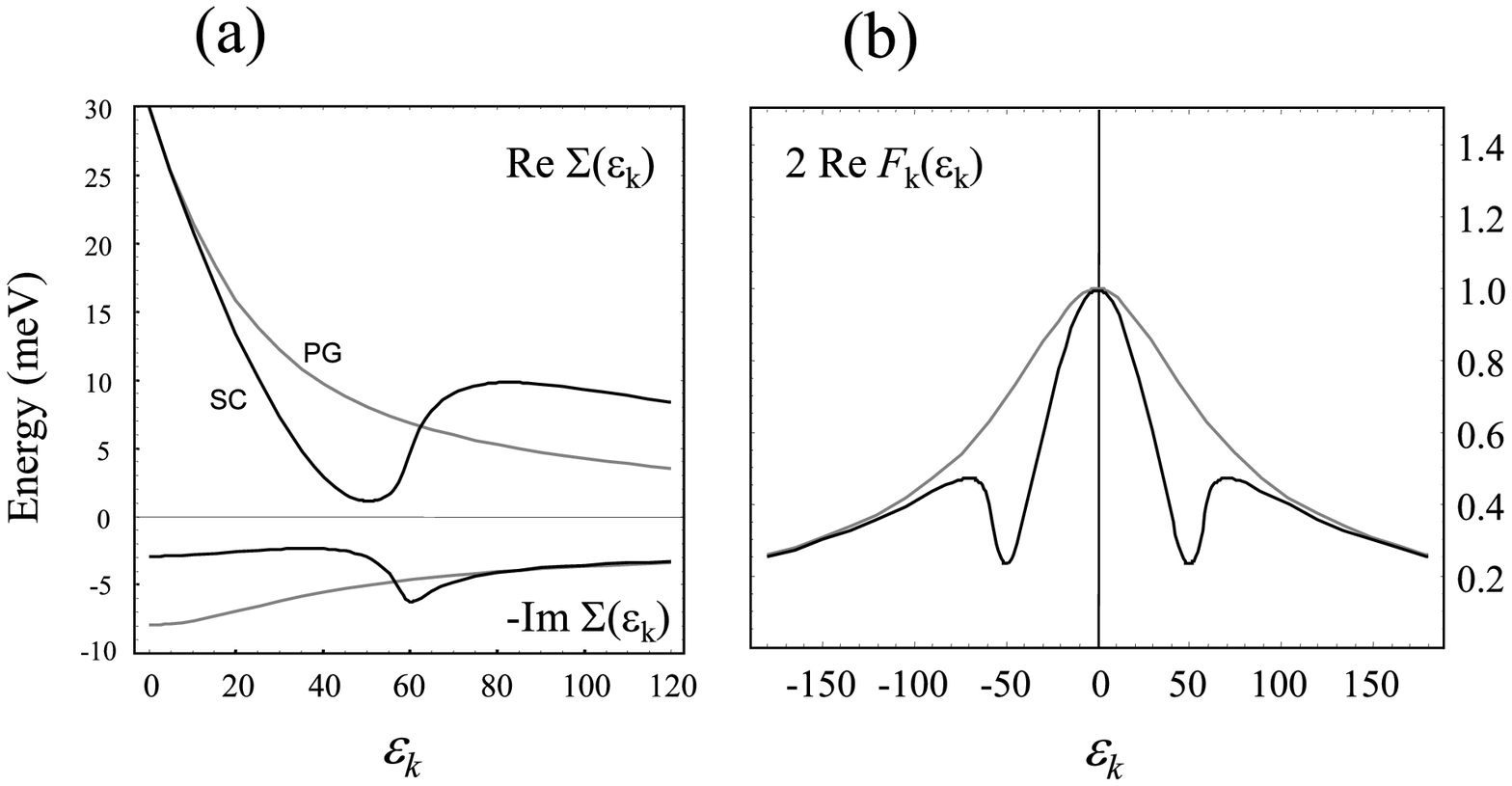}
    }
    \end{figure}

{\small Fig.\,12a. Complex self-energy plotted against
$\epsilon_k$ along $\theta=0$. Dark line - SC self-energy; gray
line - PG case.

The strong dip, where ${\rm Re}\,\Sigma(\epsilon_k)$ nearly
vanishes, is followed by a crossing-point with the PG line, then
slowly decays at higher energy. The sudden rise in ${\rm Re}
\Sigma(\epsilon_k)$, at a {\it maximum} of ${\rm Im}
\Sigma(\epsilon_k)$, is at the energy $\epsilon_k\simeq
2\sqrt{\Delta_0\,\Delta_\varphi}$, and causes the DOS dip. The
sharpest QP peaks are not at $\epsilon_k\sim0$, $E\sim\Delta_p$,
but nearer $\sim 40\,$ meV.

Fig.\,12b. Plot of the pair amplitude $2\,F_k = \Delta_k(E)/E_k$,
using the gap function (18), and compared to the BCS value, along
the anti-nodal direction. (All parameters are the same as in
Figs.\,8-10.) }

\vskip 2 mm

In conclusion, the self-energy of Fig.\,12a is yet another way of
comparing the change of states from the PG to SC cases. Indeed,
the self-energy of the SC state is smaller than the PG one for
small energies, up to the crossing point, when the relative
magnitude inverts. Our previous observation that states are
removed at large energy from the PG state, for $\epsilon_k \gg
2\sqrt{\Delta_0\,\Delta_\varphi}$, while they are added to the SC
state, for $\epsilon_k<2\sqrt{\Delta_0\,\Delta_\varphi}$, is seen
here in a different way. The strong minimum of $Re \Sigma$, near
$\epsilon_k\simeq 2\sqrt{\Delta_p\,\Delta_\varphi}$, also reveals
the condensation amplitude. The sharpness of these features is an
indication that ARPES may be a better technique to extract these
two parameters, as compared to quasiparticle tunneling.

In Fig.\,12b we plot the pair amplitude $F_k = \Delta_k/2E_k$ as a
function of $\epsilon_k$ \cite{degennes}, taking the real part.
The case where the SC gap function is used is directly compared to
the conventional pair amplitude, with a constant gap. The obvious
dip features, near $\pm$ 50 meV, are due to the same minimum
discussed previously. The investigation of this function is yet to
be done in the case of cuprates; it requires the Andreev effect or
pair tunneling, and not quasiparticle tunneling, in order to be
probed.

\vskip 2 mm

\section{V. Conclusions}


Three principal questions have been addressed\,: does the tunneling
DOS reveal the order parameter and if so, in what manner, and finally,
does it connect to the phase diagram\,? We have tried to answer them
from the heuristic point of view, i.e. by using our resonant gap
function and examining the available data. This $E$-dependant gap
produces the well-known quasiparticle peak shapes, followed by the
dip structures at higher energies. No Fermi surface anisotropy, no
van Hove singularity, no strong-coupling are needed; only simple $d$-wave
quasiparticles, with a modified spectrum, are considered.

The model contains two fundamental energy scales, whose values are
determined by the fits to the tunneling spectra\,: $\Delta_p$, the
quasiparticle peak position, and a second energy,
$\Delta_\varphi$. We found the resonance (along the anti-nodal
direction) to be at the fixed position $\Delta_0 = \Delta_p +
\Delta_\varphi$, coinciding with the total spectral gap for high
energies, i.e. $E_k \simeq \sqrt{\epsilon_k^2 + \Delta_0^2}$.
Thus, the description of the SC state depends uniquely on this new
energy scale\,: $\Delta_\varphi$.

To go a step further, we showed that these two energy scales fit the
phase diagram as a function of carrier density, $p$. While
$\Delta_p$ decreases linearly with doping, $\Delta_\varphi$ follows
$T_c$, with $\Delta_\varphi = 2.3\,k_B\,T_c$, and merges smoothly
with $\Delta_p$ at high doping. Thus the gap function scales
perfectly with the carrier density; one can then generate the QP-DOS
shape for arbitrary $p$. The two energy scales have a simple link\,:
$\Delta_\varphi \propto p\,\Delta_p$, valid in the range of the
available data. If there is a critical point, it would be at maximal
doping where both energy scales vanish.

The role of $\Delta_\varphi$ as a new condensation amplitude is
compelling. To check this, we estimated the energy changes
associated with the three states (normal, PG, SC). A simple
interpretation for each term of the constraint $\Delta_0 =
\Delta_p + \Delta_\varphi$ is, respectively\,: the total energy
gain (per pair), the energy gain due to the precursor gap (PG
state) and the condensation energy per pair $E_\varphi/N_p$.
However, the number of gapped pairs turns out to be $N_p \propto
\Delta_p/2$ and {\it not} $N_p \propto \Delta_0/2$, which is thus
in support of the preformed-pair model. For a competing order gap,
$N_p$ is to be simply taken as half the number of gapped states.
In either case our new amplitude turns out to be $\Delta_\varphi =
\sqrt{2}\,E_\varphi/N_p$. It is therefore justified, for both
scenarios, that this parameter is proportional to $T_c$, as found
independently from the fits to the data.

Since the mouvement of states in the PG to SC transition remains
subtle, i.e. not the direct opening of a second (SC) gap, we took
a closer look at the dispersion law\,$E_{\bf k}(\epsilon_k)$. We
discussed the spectral function, and the corresponding
self-energy, where the change of the quasiparticle states can be
seen more readily than in the DOS. Indeed, it allows to identify
features of the QP spectrum as a function of both $E$ and
$\epsilon_k$. One finds that in the PG state quasiparticles of
higher energy, $\epsilon_k \gg 2\sqrt{\Delta_0\,\Delta_\varphi}$,
end up as quasiparticle peaks at lower energy,
$\epsilon_k<2\sqrt{\Delta_0\,\Delta_\varphi}$, in the SC state.
The cross-over is precisely the characteristic energy where the
dip feature is seen, either in the self-energy or in the gap
function. By a different analysis than of the DOS, we found that
this mouvement of the states depends uniquely on $\Delta_\varphi$.

Our model shows that the principal change of kinetic and potential
energies (as in BCS) is associated with the opening of the
precursor gap $\Delta_{PG}\simeq\Delta_p$, at the Fermi level
(normal state to PG state transition). At the SC transition, the
quantity $E_\varphi$ is therefore linked to other degrees of
freedom not accounted for in the BCS model. This possibility is
accentuated by the resonant character of the gap function. While a
particular collective effect is perhaps responsible, a likely
candidate for the condensation energy is the phase stiffness, $J$,
in the Kosterlitz-Thouless transition. In this context, we get an
immediate relation between our parameter and $J$\,:
$\Delta_\varphi = \sqrt{2}\pi\,J$ which is then proportional to
the KT transition temperature. A relevant prediction is thus\,: $J
\propto p\, \Delta_p$, should the KT mechanism fit the phase
diagram, for a wide range of doping.

The three questions above have therefore been given a tentative
answer. One could raise some objection to introducing an {\it ad
hoc} gap function, where a microscopic theory is missing. Some
speculations concerning the physical origin of the gap function
have been given in the closing paragraphs of Sec.\,III. Our main
goal of matching the observed experimental QP-DOS, and their
characteristics as a function of the carrier density, works
without resorting to additional extraneous factors. Moreover, the
fits can now be done with a minimal set of parameters. Finally,
the two key parameters, the QP peak position $\Delta_p$, and the
coherence amplitude $\Delta_\varphi$, have been connected to
observed physical quantities.

\vskip .2 cm

{\small One of the authors (W.S.) thanks Amir Kohen \&
Th.\,Proslier, for interesting and helpful discussions.}



\end{document}